\documentclass{emulateapj}

\usepackage{amsmath}
\usepackage{amssymb}
\usepackage{amsfonts}
\usepackage{amstext}
\usepackage{graphicx}
\usepackage{fancybox}
\usepackage[usenames,dvipsnames]{color}
\usepackage{pstricks}
\usepackage{bm}
\usepackage{soul, color}
\usepackage{latexsym}
\usepackage{pifont}
\usepackage{shorttoc}
\usepackage{subfigure}
\usepackage{textcomp} 
\usepackage{float}
\usepackage{enumitem}

\usepackage{hyperref}
\usepackage{bm}
\usepackage{lipsum}
\usepackage{threeparttable}

\newcommand{\lya}{Ly$\alpha$\ }

\newcommand{\angs}{\, {\rm \AA}}

\newcommand{\no}[1]{}%per comentar espais grans

\newcommand{\myemail}{l.m.ribas@astro.uio.no}
\def\lsim{~\rlap{$<$}{\lower 1.0ex\hbox{$\sim$}}}

\def\gsim{~\rlap{$>$}{\lower 1.0ex\hbox{$\sim$}}}

\shorttitle{OMG I: Catalogs of Metal-line Absorption Doublets from High-Res Quasar Spectra}
\shortauthors{Ll. Mas-Ribas et al.}

\begin{document}

\title{Origin of Metals around Galaxies I: Catalogs of Metal-line Absorption Doublets from High-Resolution Quasar Spectra}

\author{Llu\'is Mas-Ribas\altaffilmark{1}} 
\author{Signe Riemer-S\o rensen\altaffilmark{1}}
\author{Joseph F. Hennawi\altaffilmark{2,3}} 
\author{Jordi Miralda-Escud\'e\altaffilmark{4,5}} 
\author{John M. O'Meara\altaffilmark{6}}
\author{Ignasi P\'erez-R\`afols\altaffilmark{7}}
\author{Michael T. Murphy\altaffilmark{8}}
\author{John K. Webb\altaffilmark{9}}
\altaffiltext{1}{Institute of Theoretical Astrophysics, University of Oslo,
 Postboks 1029, 0315 Oslo, Norway. {\bf \myemail}}
\altaffiltext{2}{Department of Physics, University of California, Santa 
Barbara, CA 93106, USA}
\altaffiltext{3}{Max-Planck-Institut f{\"u}r Astronomie, K{\"o}nigstuhl 17, 
D-69117 Heidelberg, Germany}
\altaffiltext{4}{Institut de Ci\`encies del Cosmos, Universitat de Barcelona (UB-IEEC),
Barcelona 08028, Catalonia.}
\altaffiltext{5}{Instituci\'o Catalana de Recerca i Estudis Avan\c{c}ats, Barcelona,
Catalonia.}
\altaffiltext{6}{Saint Michael?s College, One Winooski Park, Colchester, VT
05439, USA}
\altaffiltext{7}{Aix Marseille Universit\'e, CNRS, LAM (Laboratoire d'Astrophysique de Marseille) UMR 7326, F-13388, Marseille, France}
\altaffiltext{8}{Centre for Astrophysics and Supercomputing, Swinburne University of Technology, Hawthorn, Victoria 3122, Australia 0000-0002-7040-5498}
\altaffiltext{9}{School of Physics, University of New South Wales, Sydney NSW 2052, Australia}

%-------------------------------ABSTRACT---------------------------
\begin{abstract}

     We present the first paper of the series Origin of Metals around Galaxies (OMG) aimed to study 
the origin of the metals observed in the circumgalactic and intergalactic media. In this work we extract 
and build the catalogs of metal absorbers that will be used in future analyses, and make our results publicly 
available to the community. We design a fully automatic algorithm to search for absorption metal-line 
doublets of the species CIV, NV, SiIV and MgII in high-resolution ($R\gsim30\,000$) quasar spectra without 
human intervention, and apply it to the high-resolution and signal-to-noise ratio spectra of 690 quasars, 
observed with the UVES and HIRES instruments. 
We obtain $5\,656$ CIV doublets, $7\,919$ doublets of MgII, $2\,258$ of SiIV, and 239 of NV, constituting the 
largest high-resolution metal-doublet samples to date, and estimate the dependence of their completeness and 
purity on various doublet parameters such as equivalent width and redshift, using real and artificial 
quasar spectra. The catalogs include doublets with rest-frame line equivalent widths down 
to a few ${\rm m\angs}$, all detected at a significance above 3$\sigma$, and covering the redshifts between 
$1<z \lesssim 5$, properties that make them useful for a wide range of chemical evolution studies.

\end{abstract}

%-------------------------------INTRODUCTION----------------------------------------------------------------
\section{Introduction}

   Metals are precise probes for revealing the physical properties of the medium that they inhabit, ranging  
from small regions within galaxies out to the largest scales of the intergalactic medium (IGM), and from 
the present day back to the redshifts of cosmic reionization \citep[e.g.,][]{Tytler1995,Songaila1996,Pettini2001}. 

   Despite their importance, our knowledge on the origin of metals is restricted to those detected in 
massive galaxies and their immediate surroundings, where local star formation and feedback 
processes create and distribute them, respectively. It remains unclear whether the main polluting 
mechanism of the IGM are powerful feedback processes from distant massive and bright galaxies, 
or the in situ enrichment by (currently) undetectable small-galaxy populations embedded in such a 
medium \citep{Pettini2003,Bertone2005,Bouche2006,Bouche2007,Pratt2017}. 
\cite{Adelberger2003} proposed that galactic superwinds from massive galaxies 
were the mechanism responsible for the abundance of CIV absorption systems detected around $z=3$ 
Lyman Break Galaxies, but \cite{Porciani2005} argued that an early metal pollution driven by dwarf galaxies 
at $6<z<12$ was a more likely scenario. The latter origin was also supported by the observations of 
\cite{Simcoe2011}, suggesting that half of the intergalactic metals observed at $z\sim2.4$ were produced  
at higher redshifts, between $z\sim4.3$ and $z\sim2.4$, consistent with the enrichment by small 
galaxies (with dark matter 
halos of masses $M_{\rm h}\lesssim 10^{11}\,{\rm M_{\odot}}$) found in the simulations by 
\cite{Wiersma2010}. More recent simulations by \cite{Oppenheimer2012} also favored an IGM enriched 
by similar low-mass galaxies through supernova feedback, arguing that metals 
produced in massive systems might not be able to escape the galactic environment \citep[see 
also][]{Hayward2017}.

    A powerful method to probe the origin of metals around galaxies and in the IGM is the study of their 
clustering signature \citep[e.g.,][]{Adelberger2005,Scannapieco2006,Wild2008,Martin2010,
PerezRafols2014}. In detail, the clustering of cosmic structure follows that of the 
underlying dark-matter density field as $\xi (r) \simeq b^2\, \xi_{\rm DM}(r)$, where $\xi(r)$ and 
$\xi_{\rm DM}(r)$ denote the distance dependent clustering (i.e., the two-point correlation function) 
of structure and dark matter, respectively, and the term $b$ is the 
so called {\it bias factor} \citep{Sheth1999,Tinker2010}. This bias depends on the mass of the dark matter halo 
hosting the structure of interest \citep{Zehavi2005}, the redshift, and the spatial scale, although in the linear 
regime (large scales) the scale dependence is small \citep{Peebles1980,Peacock1996}. One 
expects the metals and the galaxy population that produces them to inhabit the same dark matter halos 
and show the same clustering, and therefore, to also have the same bias factor, although the evolution of 
a population of halos of fixed mass modifies their bias factors if the metals are observed a long time after 
they were produced. In conclusion, the measurement of 
the bias factor of the intergalactic metals can be used as an indicator of their progenitor galaxies.

   Shedding light on the sources responsible for enriching the IGM through the  measurement of the metal bias 
factor is the main goal of our project Origin of Metals around Galaxies (OMG). We aim to obtain the metal bias 
by spatially cross-correlating the intergalactic metals found in a specifically built sample of quasar spectra, 
with the Lyman-alpha (Ly$\alpha$) forest from the quasar spectra in the DR12 catalog 
\citep[DR12Q,][]{Paris2016} of the BOSS/SDSS-III Collaboration \citep{Eisenstein2011,Dawson2013,
Alam2015}. In principle, we could calculate the auto-correlation function of the metals alone, but the number 
of detectable metal absorbers 
is small (a few thousands) and this would result in weak constraints. Since the number of useful 
pixels in the BOSS quasar spectra denoting the \lya forest is large \citep[$\sim 27$ millions;][]{Busca2013}, 
the cross-correlation between the two tracers is a better choice. Overall, this cross-correlation will equal 
$\xi_{\rm metal-Ly\alpha}\simeq b_{\rm metal}\,b_{\rm Ly\alpha}\,\xi_{\rm DM}$, which is the product of the 
clustering  of the underlying dark matter density field, the bias factor of the \lya forest $b_{\rm Ly\alpha}$, and 
the bias factor of the population of metal absorbers. The first term can be calculated from the theory of 
structure formation within the Lambda 
Cold Dark Matter ($\Lambda$CDM) cosmological model \citep[e.g.,][]{Smith2003}, and the \lya forest bias 
has been constrained in the recent works by \cite{Blomqvist2015}, \cite{Delubac2015}, and 
\cite{Bautista2017}, so that the bias factor of the metal population can be determined from the clustering. 

    For the calculation of the IGM metal bias, we will use weak CIV$\,\lambda\lambda$1548,1550 
metal absorption doublets detected in high-resolution quasar spectra, motivated by the following 
aspects: ({\it i}) we need the largest possible number of metal systems spatially overlapping with 
the \lya forest data in order to reduce the statistical uncertainties of the measurements and obtain tight 
constraints on the bias factor. CIV is the optimal 
ion for this purpose due to the large cosmic abundance of carbon, which typically results in the detection 
of several CIV doublets in one quasar spectrum, the exact number depending on the signal-to-noise ratio and 
quasar redshift. Furthermore, the doublet nature of the absorption systems facilitates their identification, and 
the value of the CIV rest-frame wavelength enables covering a broad redshift range when 
the doublets are identified in ultraviolet spectra. 
({\it ii}) the requirement of weak absorption systems arises from the fact that the strength of the absorption 
lines is broadly correlated with the metal content \citep[e.g.,][]{Masribas2016c} and, in turn, 
the mass of the host medium. We expect strong absorbers to trace massive galaxies instead 
of the IGM. In \cite{Masribas2016c}, we found a mean rest-frame equivalent width 
of $\langle W_r\rangle \sim0.43\,\angs$ for the CIV$\,\lambda1548$ line of the carbon doublet by 
stacking the spectra of $\sim 27\,000$ Damped Lyman Alpha systems \citep[DLAs; see][for a review]{Wolfe2005}, 
a type of objects associated to galaxies with dark matter halo masses in the range $8.7 
\lesssim \log (M_{\rm h}/{\rm M_{\odot}}) \lesssim 11.8$ 
\citep[][see also \citealt{Fontribera2012}]{Perezrafols2017}. Furthermore, \cite{Vikas2013} argued 
that CIV systems with $W_r>0.28\,\angs$ 
and up to $W_r\sim5\,\angs$ inhabit dark matter halos with masses of $\log (M_{\rm h}/{\rm M_{\odot}}) 
\gtrsim 11.3-13.4$, and \cite{Gontcho2017} recently found indications suggesting that weaker absorbers may 
represent smaller halos when comparing their findings with those by Vikas et al\footnote{The aforementioned 
findings are also supported by studies of MgII absorbers at redshifts $z\lesssim2$; \cite{Mathes2017} argued 
that weak and strong MgII absorbers represent different objects given their different evolution with redshift, 
and \cite{Kacprzak2011} suggested 
that weak MgII absorbers trace cold filaments that are being accreted onto the galaxies.}. In view of these results, 
it seems plausible to consider CIV doublets with $W_r \gtrsim 0.3\,\angs$ to be tracers of massive galaxies 
and not desirable for our purposes, although the exact threshold should be revisited when doing the cross-correlation. 
({\it iii}) the highest possible purity ($>90\%$) is needed for obtaining a reliable bias factor value because each 
false-positive detection contributes to the bias calculation with a null value that, overall, results in an   
underestimation of the true result. Summarizing, we need to search for a large number of  CIV doublets with 
small equivalent widths, while ensuring the minimum number of false-positive detections, which is only possible 
with spectra of very high resolution and signal-to-noise ratio. This metal search is the purpose of this first paper of the 
OMG series.   

%------------------------------ Observational requirements ----------------------------------------------------------------
\subsection{Observational Requirements}

   The first aspect of our calculations is obtaining a large enough sample of high-resolution quasar 
spectra. In \cite{Perezrafols2017}, we recently cross-correlated $\sim 14\,000$ DLAs (dataset A) 
with the Lyman alpha forest sample by \cite{Busca2013}, which resulted 
in one-sigma uncertainties of around $10\%$ for a bias factor value of $b_{\rm DLA}\sim2$. Assuming a 
similar bias factor for CIV and DLAs, and that we can detect ten CIV systems in each quasar 
spectrum on average, consistent with our results presented below, obtaining similar constraints 
would require the search of $1\,400$ high-resolution and signal-to-noise ratio quasar spectra. In practice, 
however, not all the 
detected CIV systems will be weak, which results in the need for even a larger number of spectra. 
As we will show below, our  sample consists of 690 quasars with high-resolution ($R\gsim 30\,000$) spectra 
and median signal-to-noise ratios above 10. 
Despite this number of spectra being lower than our rough estimate, it represents the largest compilation of quasar 
spectra of this resolution to date, followed by the partially overlapping sample of 602 quasars 
by \cite{Mathes2017}, who investigated properties of MgII absorbers. 
The use of our quasar sample will constitute a strong improvement compared to previous works, which were 
only able to use a large number of spectra for the case of low-resolution ( $\sim 2000$, typical of the BOSS 
quasar spectra) and low signal-to-noise ratio, and, therefore, could only identify the strong absorbers that probe 
the vicinity of massive galaxies.

     Since our methodology for searching quasar spectra can be applied to metal 
doublets of different species  
in a similar fashion, we perform calculations to obtain doublet catalogs of CIV, SiIV, NV, and MgII, and 
make these catalogs publicly available at \url{https://github.com/lluism/OMG}, 
which will enable a large number of additional metal studies by the community. We emphasize that, 
given our purpose, we will pay special attention to building catalogs with high purity, and will be 
less concerned about completeness.

  This paper is structured as follows. In \S~\ref{sec:data} we present the quasar data and, in 
\S~\ref{sec:code}, we detail the search code designed to find the doublets. The capabilities of the 
code are tested in \S~\ref{sec:test}, where we estimate the completeness and purity expected from 
the search. The resulting metal-doublet catalogs are presented in \S~\ref{sec:catalogs}, before  
concluding in \S~\ref{sec:conclusion}.

%-------------------------------------- OBSERVATIONAL DATA  ----------------------
\section{Quasar Spectra}\label{sec:data}

\begin{figure}\center      % QSO SNR and Z
\includegraphics[width=0.47\textwidth]{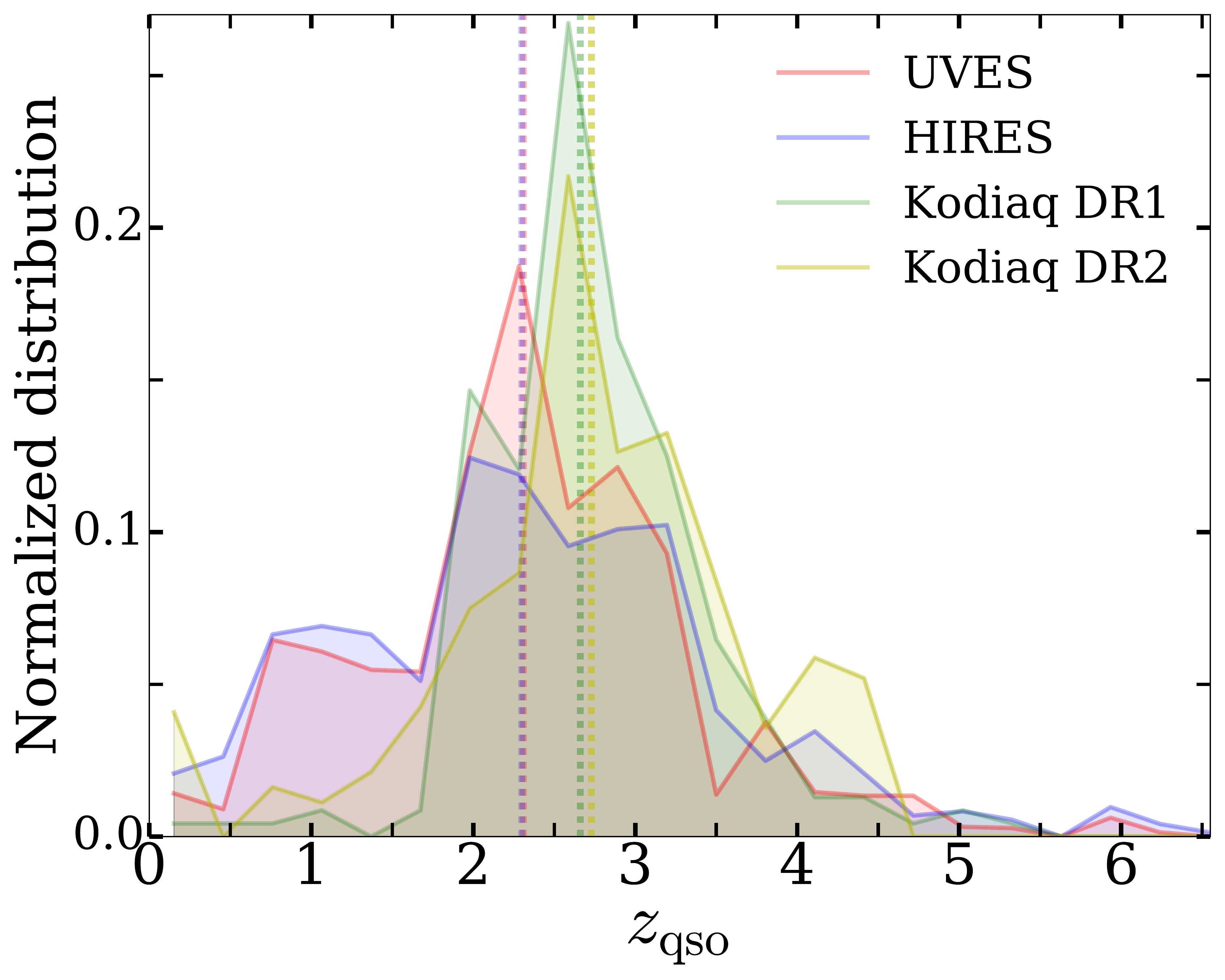}
\includegraphics[width=0.47\textwidth]{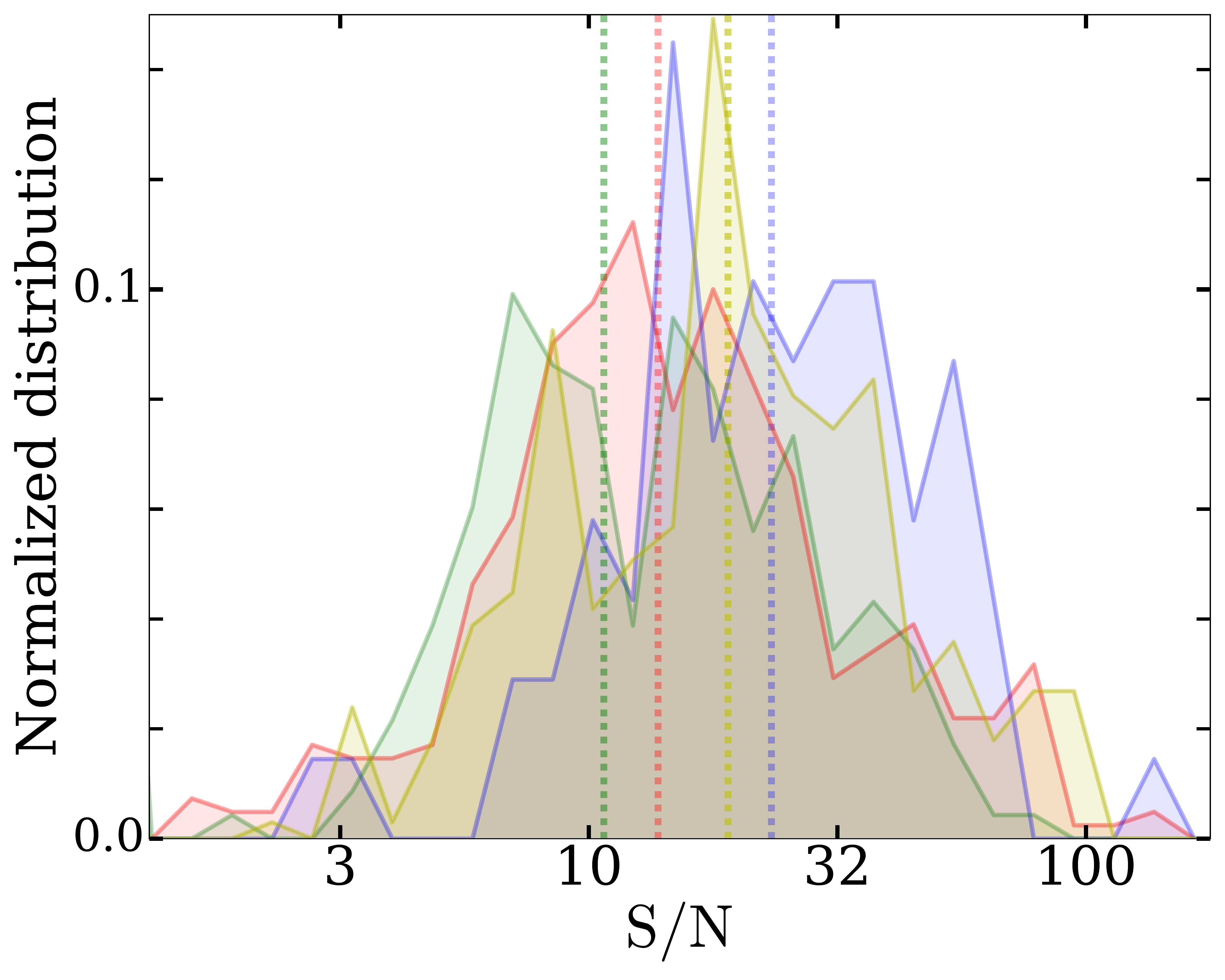}
\caption{{\it Upper panel:} Redshift distributions for the four quasar samples in our study. The {\it vertical 
dashed lines} denote the median of each sample. {\it Lower panel:} Distributions for the signal-to-noise 
(S/N) ratio of each quasar sample. The {\it vertical dashed lines} represent the median of the samples, 
calculated considering the whole wavelength range covered by each spectrum.} 
\label{fig:sample}
\end{figure}

\begin{table}\center
	\begin{center}
	\caption{Quasar samples}\label{ta:qso}
	\begin{threeparttable}
		\begin{tabular}{cccc} 
		\hline
		\hline	
		Sample                              & S/N\tnote{\,a}       &$\bar z$     &Num. of quasars    \\ 
		\hline  
		UVES	          	&13.8		&$2.3$	&$414$       	\\  
		HIRES                	&23.3		&$2.3$	 &$71$      	 \\  
		Kodiaq DR1       	&10.7		 &$2.7$	&$170$      	\\  	
		Kodiaq DR2       	&19.2		&$2.7$	&$130$      	\\  	
		\hline
		\end{tabular}
		\begin{tablenotes}
			\item[a] Median S/N ratio considering the entire wavelength range of the spectra.
		\end{tablenotes}		
	\end{threeparttable}
	\end{center}
\end{table}

   Our data is composed of four high-resolution quasar spectra datasets: we use the continuum 
normalized spectra of the 170 quasars in the first public data release of the Keck  Observatory  Database of 
Ionized  Absorption  toward  Quasars  \citep[KODIAQ DR1;][]{Omeara2015}, and those of the 130 additional 
quasars incorporated in the second public data release \citep[KODIAQ DR2;][]{Omeara2017}\footnote{\url{https://koa.ipac.caltech.edu/workspace/TMP_AqLrga_13996/kodiaq13996.html}}. 
We also consider $414$ and $71$ quasars with continuum normalized spectra from observations with the 
Ultraviolet and Visual Echelle Spectrograph \citep[UVES;][]{Dekker2000} at the Very Large Telescope (VLT), 
and the High Resolution 
Echelle Spectrometer \citep[HIRES;][]{Vogt1994} at Keck, respectively, \citep[][Murphy et al. 2018, in 
prep]{Murphy2003,King2012}. Table \ref{ta:qso} indicates the number of quasars for each sample, their 
median quasar redshift, and the median signal-to-noise ratio of all the spectra considering their entire 
wavelength range. The {\it upper} and {\it lower panels} of Figure \ref{fig:sample} display the quasar redshift 
and spectral signal-to-noise ratio distributions, respectively, for the four samples, and the {\it dashed vertical 
lines} denote the median of each distribution.

     The Kodiaq and HIRES samples were both built from studies using the same instrument (i.e., HIRES 
spectrograph), but we name the samples in this way because they contain different spectra. Some 
quasars are observed and contained in more than one sample, but we consider them all for the search,  
because different observations cover different wavelength ranges, and some parts 
of the spectrum with no detection or high noise in one sample are clearly detectable in another one. 
In practice, there are 690 different quasars in the samples. We will 
look for repeated doublets after the search, and will keep those with highest significance in the 
detection. In all cases, we only use the data outside the \lya forest, i.e., redward the \lya emission line of the 
quasars.

\begin{figure*}\center      % SPEC
\includegraphics[width=1\textwidth]{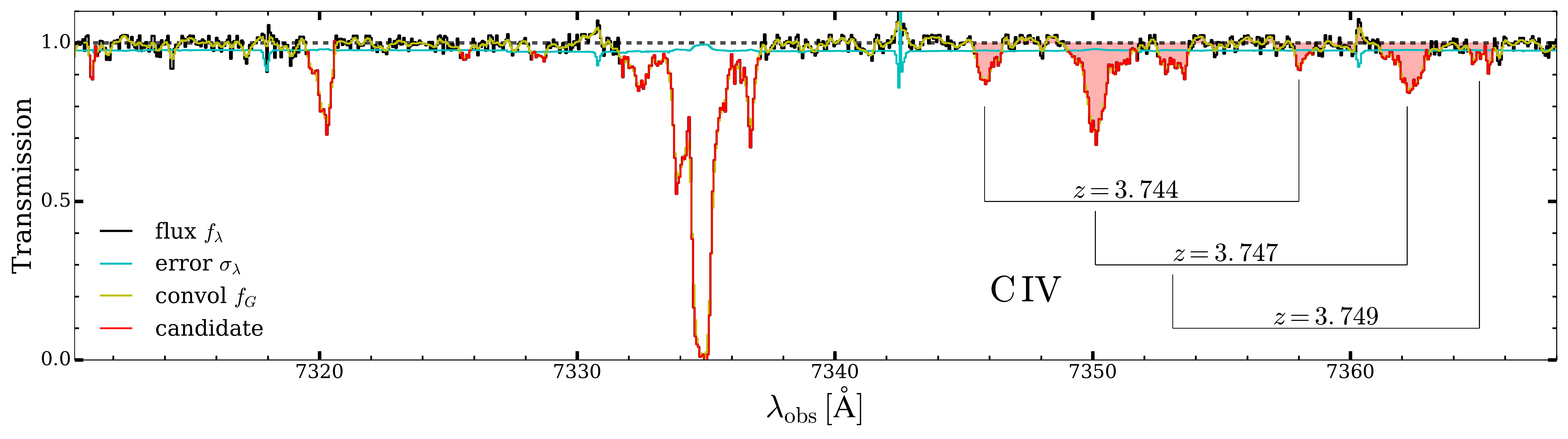}
\caption{Wavelength range of a quasar spectrum in the UVES sample. The {\it black line} 
denotes the normalized spectrum, and the {\it cyan line} is the value 1-$\sigma_{\lambda}$ 
at every pixel. The {\it yellow  
line} denotes the gaussian-convolved flux, and the {\it red lines} the candidate absorption features. 
The regions pertaining to a CIV doublet, between $\sim7345\,{\rm \angs}$ and ${\sim 7366\,{\rm 
\angs}}$, are represented as {\it shaded red areas}, and listed as three doublets in our resulting catalog,  
although they may belong to the same absorption system. The {\it horizontal dashed} line marks the position 
of the unabsorbed continuum, i.e. a transmission of 100\%. } 
\label{fig:spec}
\end{figure*}

%-------------------------------------- QUASAR CONTINUUM  ----------------------
\subsection{Quasar Continuum}\label{sec:continuum}

   The continuum (unabsorbed) spectral energy distribution (SED) of each quasar is used 
to obtain its normalized spectrum as 
\begin{align}
f_{\lambda}&=\frac{F_{\lambda}}{C_{\lambda}} ~ ,\\\nonumber
\sigma_{\lambda}&=\frac{E_{\lambda}}{C_{\lambda}} ~,
\end{align}
where $F_{\lambda}$, $E_{\lambda}$ and $C_{\lambda}$ are the observed quasar flux, 
uncertainty, and continuum at a given wavelength (pixel), respectively, and $f_{\lambda}$ 
and $\sigma_{\lambda}$ are the normalized flux and uncertainty. 

   For the KODIAQ data, we use the continua provided along with the observed spectra and 
computed as described in \cite{Omeara2015}. Briefly, the spectra are continuum-fitted by 
hand, using Legendre polynomials of order generally between 4 and 12, depending on the 
specific characteristics of every spectrum. The continua for the UVES and HIRES spectra 
were obtained by fitting spline curves to regions that are free from absorption 
features (as described in \citealt{Bagdonaite2014} and \citealt{Signe2015}). 
For the UVES spectra, a 6 to 8 order Chebyshev polynomial was generally used.  
Because our analysis uses only continua outside the Lya forest, where absorption lines are rare, we 
expect the impact of continuum uncertainty to be negligible, but we perform a more quantitative test in 
\S~\ref{sec:testcont}.

%-------------------------------------- CODE ----------------------
\section{Blind Doublet Search Code}\label{sec:code}

   We design a code to detect metal absorption doublets in quasar spectra, 
based on the public code by \cite{Cooksey2008,Cooksey2010}\footnote{\url{http://guavanator.uhh.hawaii.edu/
~kcooksey/igmabsorbers.html}}$^,$\footnote{\url{http://cvs.ucolick.org/viewcvs.cgi/xidl/HST/CIV/?cvsroot=
Prochaska}}, which has proven to be efficient for this purpose \citep[e.g.,][]
{Cooksey2011,Cooksey2013,Seyffert2013}. 
Contrary to the Cooksey code, our code is built to be fully automatic, without the need 
for human intervention, which allows searching a large number of spectra in 
a short time. 
%We tune the parameters in the code to obtain metal doublet catalogs 
%with the maximum possible purity ($>$), i.e., a low fraction of false detections, which will be the important 
%requirement for our future studies, at the cost of lower completeness.

   In \S~\ref{sec:search} below, we describe the code focusing on the procedure for 
the search and identification of CIV$\,\lambda\lambda$1548,1550 doublets, although the 
code also searches for the additional doublets SiIV$\,\lambda\lambda$1393,1402, 
NV$\,\lambda\lambda$1238,1242 and MgII$\,\lambda\lambda$2796,2803, simultaneously. 
In \S~\ref{sec:accept}, the requirements that the doublet candidates have to 
accomplish in order to be considered real and included in the final catalogs are presented.

%--------------------------------------  SEARCH  ----------------------
\subsection{Automatic Metal Doublet Search}\label{sec:search}

   The code starts by correcting possible outliers in the spectrum: the normalized pixel fluxes are limited  
to the values $-\sigma_{\lambda} \le f_{\lambda} \le 1+\sigma_{\lambda}$. Pixels with flux values outside 
this range are reset to the upper/lower limit. Additionally, 
we limit the flux between $-1 \le f_{\lambda} \le 2$ to account for unreasonable flux values usually 
associated with extremely 
large uncertainties. 

     After these corrections, we convolve the spectrum using a Gaussian kernel with a FWHM of three times 
the FWHM of the instrument, and group adjacent pixels whose Gaussian-convolved flux is 
$f_{G} \le 1-\sigma_{\lambda}$, also including the pixels outside this group that reside within one kernel 
sigma from both group sides. Only groups with a minimum number of pixels between three 
and five, depending on the sample, are considered. Each of these groups then represents an 
absorption candidate within 
a range defined by a minimum and maximum wavelength, $\lambda_l$ and $\lambda_h$, 
respectively. Figure \ref{fig:spec} displays an example of the search described above 
for an arbitrary wavelength interval and a spectrum from the UVES sample. The {\it black line} denotes the 
normalized flux and the {\it cyan line} the value 1-uncertainty at every pixel (1-$\sigma_{\lambda}$). 
The {\it yellow line} represents the Gaussian convolved flux, and the {\it red line} denotes the candidate 
absorption features whose convolved flux resides below the 1-$\sigma_{\lambda}$ line. 

   Under the assumption that these candidates are ${\rm CIV\,\lambda1548}$ absorption lines (or the 
shortest wavelength line of the doublet for other species), we  
calculate their redshift and equivalent width as follows.  
The absorption redshift is calculated as \citep{Cooksey2010}
\begin{equation}
1+z_{\rm abs}=\frac{\sum \limits_{\lambda_l}^{\lambda_h} \lambda_i \ln\left(1\over f_{\lambda i}\right)}{\lambda_r \sum \limits_{\lambda_l}^{\lambda_h} \ln\left(1\over f_{\lambda i}\right)} ~,
\end{equation}
where $\lambda_r$ is the rest-frame wavelength of the metal line, and $\lambda_i$ is 
the observed wavelength of the pixel $i$. The logarithmic term  
denotes the pixel optical depth and acts as a weight, i.e., pixels with large optical depth 
will contribute more to the redshift calculation. For the calculation of the weights, we limit 
the minimum flux to $f_{\lambda} \ge (0.2 \sigma_{\lambda} > 0.05)$ to avoid extremely 
large values of the optical depth where the line saturates \citep{Cooksey2010}. 

   The rest-frame equivalent width and its uncertainty from error propagation equate  
\begin{eqnarray}
W_r &=& {1\over(1+z_{\rm abs})} \sum \limits_{\lambda_l}^{\lambda_h} (1-f_{\lambda i}) \delta \lambda_i ~,\\  \nonumber
\sigma_{W_r}^2 &=& {1\over(1+z_{\rm abs})^2} \sum \limits_{\lambda_l}^{\lambda_h} \sigma_{\lambda i}^2 {\delta \lambda_i^2} ~,
\end{eqnarray}
where $\delta \lambda_i$ is the wavelength interval between pixels.

   We use the redshift measurements to calculate the expected position of the 
associated ${\rm CIV\,\lambda1550}$ lines (or the higher wavelength lines of the other doublets). 
In practice, the actual position of the 
associated lines fluctuates around the calculated value due to the effects of blends with 
other lines and noise, and to the fact that we imposed a flux limit for the weights in the redshift 
calculation. A match is accepted when the 
velocity offset between the observed and calculated positions of the associated lines 
is $\lvert \delta_v \rvert \le 7\, {\rm km\,s^{-1}}$. This value, as those of other parameters, is set to 
maximize the number of detections while avoiding false positive 
detections (see \S~\ref{sec:test}). 

   Our code also computes the column density of each 
feature as in \cite{Cooksey2008}, using the Apparent Optical Depth (AOD) method by 
\cite{Savage1991}. However, we do not present these measurements because this method is 
valid only when the absorption lines are unsaturated \citep[][although see also \citealt{Cooksey2008}]
{Fox2005}, which occurs only for $\sim 50-60\%$ of the doublets.

%-------------------------------------- ACCEPTANCE  ----------------------
\subsection{Doublet Acceptance Criteria}\label{sec:accept}

   All the previous candidate doublets are finally considered real if they accomplish all the 
following requirements (with the corresponding values for each species): 

\begin{enumerate}
\item The redshift of the metal absorption lines is below that of the quasar, 
$z_{\rm abs} \le z_{\rm em}$, where $z_{\rm em}$ is the redshift of the quasar. 

\item Both lines in the doublet reside outside the \lya forest region. 

\item The equivalent width ratio between the doublet lines, 
$R_W$, is within the range $1-\sigma_{R_W} \le R_W \le 2 + \sigma_{R_W}$, 
where $\sigma_{R_W}$ is the uncertainty of the ratio. The equivalent width ratio and 
uncertainty from error propagation are calculated as \citep{Cooksey2010}
\begin{align}
R_W = \frac{W_{r,1548}}{W_{r,1550}}~, \hspace{4.2 cm} \\ 
\sigma_{R_W}^2 = R_W^2 \left[ \left(\frac{\sigma_{W_{r,1548}}}{W_{r,1548}}\right)^2+\left(\frac{\sigma_{W_{r,1550}}}{W_{r,1550}}\right)^2\right] ~. \nonumber
\end{align} 
These limits denote the extreme cases of completely saturated, $R_W = 1$, 
and unsaturated lines, $R_W = 2$. 

\item The significance of the detection for the strongest doublet line is $W/\sigma_{W} \ge 3$.

\end{enumerate}

    The {\it red shaded regions} of Figure \ref{fig:spec} illustrate three separate CIV doublet features, a 
central component at $z\sim3.747$, plus two more within a velocity offset ranging between 
$100-200$  km ${\rm s^{-1}}$. We include these features in the final catalog as three different doublets, 
thus allowing to either group and treat them as a single absorption system or individually. There is 
another CIV doublet at the positions $\approx 7320\,{\rm \AA}$ and $\approx 7332\,{\rm \AA}$ 
($z\sim3.728$) not identified by the code. The strong absorber centered at $\approx 7335\,{\rm 
\AA}$ is blended enough with the rightmost wavelength line of the doublet so that the code considers them as a 
single feature. This results in the determination of a higher (incorrect) redshift for the redder doublet 
feature compared to that of the blue one, and, especially, fails the requirement of the equivalent width 
ratio (condition number 3 above).

%-------------------------------------- CODE TESTS ---------------------------------------
\section{Code and data validation}\label{sec:test}

  We present below a series of tests to assess the capabilities of our search code and 
the validity of our results. 
%In \S~\ref{sec:kathy}, we compare the results from our code 
%with those from using the code by \cite{Cooksey2008}. 
The impact of false positive detections  on the purity is assessed 
with the real spectra in \S~\ref{sec:art}, and we make 
use of mock spectra to further test the completeness in \S~\ref{sec:mocktest}, and 
 the effects of the quasar continuum placement in \S~\ref{sec:testcont}.

\subsection{Searching for False Positives in Real Spectra}\label{sec:art}

   We run our search code on the four quasar samples setting now the theoretical separation between 
the two lines of the doublets to a slightly larger value than the true one. This implies that all the doublets 
identified by the code in this case will be false detections, and will provide a robust estimation for the purity 
of the results. 

     We set the distances between the lines of the doublet to values a few Angstrom larger than the theoretical 
ones, the exact value depending on the species, and avoiding coincidences of these new separations with 
the values for doublets of other species. Because the separation in wavelength between lines  
changes with redshift, it is possible that for a specific redshift the separation of one species matches the value 
of the separation of another species at a different redshift. However, this also happens in the real case, and so 
we enable this possibility in the analysis. We have tested that changing the separation from a few to a 
couple of tens of Angstrom produces negligible differences in the test results.

\begin{figure*}\center      % PURITY PLOT 
\includegraphics[width=1\textwidth]{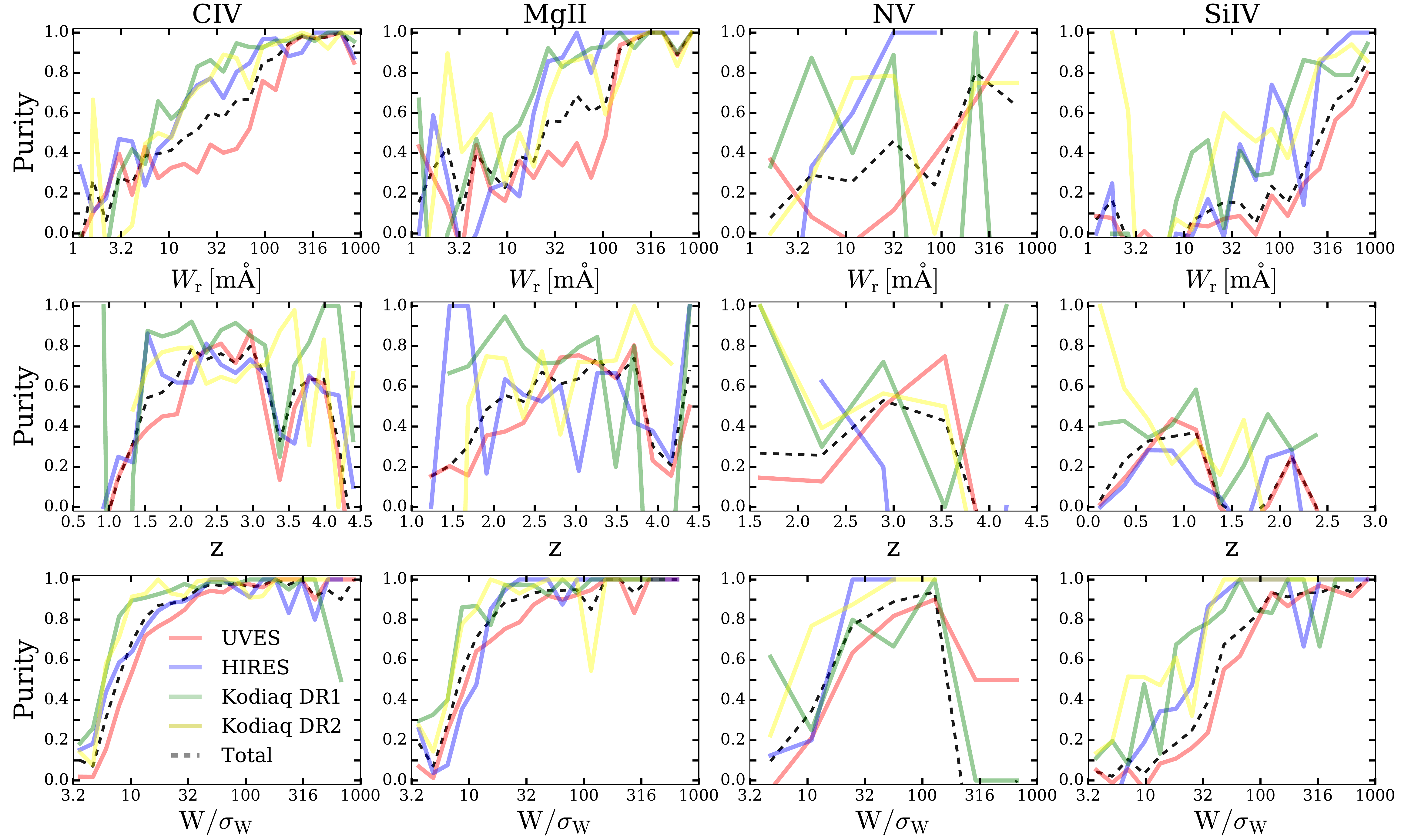}
\caption{Purity results from the search of false doublets and comparison with the total number of detections 
in the four quasar samples. The results are presented for the four 
species (from left to right panels) and equivalent width, redshift, and significance of the detection (from top 
to bottom panels), using the values of the {\bf lower wavelength} lines in the doublets. The {\it colored solid lines} denote each 
of the four samples, and the {\it dashed black line} 
represents the median of the four. The number of bins ranges between $15-25$ for visualization purposes, and 
the scales are logarithmic for the cases of the equivalent widths ({\it top panels}) and significance ({\it bottom 
panels}), and linear for the redshifts ({\it middle panels}). The best indicator for the purity is the 
significance of the detection ({\it bottom panels}), showing purities above $90\%$ at $W/\sigma_W\sim32$ 
for CIV and MgII.} 
\label{fig:purity}
\end{figure*}

   After obtaining the catalogs of false positives, we compute the purity as    
$1- {\rm false\,/\,real}$, where {\it false} and {\it real} refer to the number of false detections, and detections in the 
real search (with the true doublet line separation), respectively.  Figure \ref{fig:purity} shows the results for every 
species (from left to right panels), binned in equivalent width, redshift, and 
significance of the detection (from top 
to bottom panels), using the values of the {\bf lower wavelength} lines in the doublets. The {\it colored solid lines} denote each 
of the four samples, and the {\it dashed black line} 
represents the median of the four. 
%The number of bins ranges between $15-25$ for visualization purposes, and 
%the scales are logarithmic for the cases of the equivalent widths ({\it top panels}) and significance ({\it bottom 
%panels}), and linear for the redshifts ({\it middle panels}). 
The {\it middle panels} indicate that there is no correlation 
between the purity and the redshift of the absorbers, whilst the {\it upper panels} show an increase of purity 
with equivalent width. The equivalent widths where a purity above $90\%$ is reached varies strongly with 
the species and the sample. The significance of the detection 
({\it bottom panels}) appears as the most robust observable since the trends for the different samples present 
a narrow scatter (dispersion) around the median value. For CIV and MgII, a median purity of 
$>90\%$ is reached at $W/\sigma_W\sim32$, and at $W/\sigma_W\sim10$ is reduced to $\sim60\%$. 
For SiIV, a $W/\sigma_W\gtrsim100$ is required to reach a median purity above $90\%$, similar to the case 
of NV, although in the latter case the number of detections is too small to establish firm conclusions.

\subsection{Searching Realistic Mock Spectra}\label{sec:mocktest}

   We perform tests creating and analysing mock spectra that precisely reproduce the 
spectra from our different real-quasar data sets. 
%In \S~\ref{sec:mock}, we detail how the mock 
%spectra are built, and in \S~\ref{sec:mocksearch}, we search them to estimate the completeness 
%and purity of our results. We estimate the level of contamination from the noise in the spectra in 
%\S~\ref{sec:noise}.

%-------------------------------------- mock CREATION  ---------------------------------------
\subsubsection{Generation of Mock Spectra}\label{sec:mock}

To create mock spectra, we use the publicly available code \texttt{QSOSim10}\footnote{
\url{https://github.com/vincentdumont/qsosim}} \citep{Qsosim10}, designed to 
simulate quasar spectra from the 10th Data Release quasar catalog of the BOSS survey 
\citep[DR10Q;][]{Paris2014}. Briefly, the code 
computes a quasar continuum taking the redshift, spectral slope and magnitude parameters from the 
DR10 quasar catalog for every object, and adds up to 59 broad emission lines to it. 
The \lya forest is then created using statistical models for the redshift and column density 
distributions of neutral hydrogen, which are derived from studies of high-resolution spectra. Individual 
metal absorption lines associated to the neutral hydrogen absorption systems are included by means of  
the photoionization code \texttt{CLOUDY} v13.03 \citep{Ferland2013}. In detail, for every hydrogen absorber  
at a given redshift and hydrogen column density above $N_{\rm HI}\ge 10^{15}\, 
{\rm cm^{-2}}$, the code reads its average metallicity inferred from 
absorption spectroscopy studies and introduces these three parameters into \texttt{CLOUDY}. 
Considering the radiation background at the corresponding redshift, \texttt{CLOUDY} yields 
then the metal column densities 
of 30 atomic species and about 500 ionization states. Finally, \texttt{QSOSim10} creates absorption lines of these  
column densities with a line width sampled from a normal distribution centered at 6 km ${\rm s^{-1}}$ and 
standard deviation of 1.5 km ${\rm s^{-1}}$. Blends within different lines are allowed but the individual 
lines are created using Voigt profiles. The resulting mock spectrum 
is finally convolved accounting for the BOSS spectrograph resolution, and the noise from the 
observations and instrumentation is included. We refer the reader to \cite{Qsosim10} for more detailed 
descriptions of the code and comparisons with the real BOSS spectra.

\begin{table*}\center
	\begin{center}
	\caption{Mock spectra}\label{ta:mock}
	\begin{threeparttable}
		\begin{tabular}{ccccccccccccc} 
		\hline
		\hline	
		Sample                              &$R$       & S/N\tnote{\,a}            &Num. of spectra    & \multicolumn{4}{c}{Purity all ($\%$)}   &$\, $	&\multicolumn{4}{c}{Purity $W_r>10\,{\rm m\angs}$ ($\%$) } \\ 
							 									& & & &CIV &SiIV &NV &MgII 	&$\, $	&CIV&SiIV&NV&MgII \\
		\hline  
		mock ${\rm HIres-HIsnr}$	       &$100\,000$     	&109			&$1\,000$       	&$95$ &$97$ &$87$ &$89$    		&$\, $	&$100$ &$100$ &$99$ &$99$\\  
		mock ${\rm HIres-MDsnr}$            &$100\,000$      &43			        &$1\,000$      	&$97$ &$98$ &$89$ &$81$	&$\, $	&$100$ &$99$ &$100$ &$93$ \\  
		mock ${\rm MDres-LOsnr}$         &$50\,000$     	&7			&$1\,000$      	&$97$ &$100$ &$94$ &$85$	&$\, $	&$98$ &$100$ &$99$ &$87$\\  		
		\hline
		\end{tabular}
		\begin{tablenotes}
			\item[a] Median S/N ratio considering the entire wavelength range of the spectra.
		\end{tablenotes}		
	\end{threeparttable}
	\end{center}
\end{table*}

\begin{figure}\center      % SNR mock  
\includegraphics[width=0.47\textwidth]{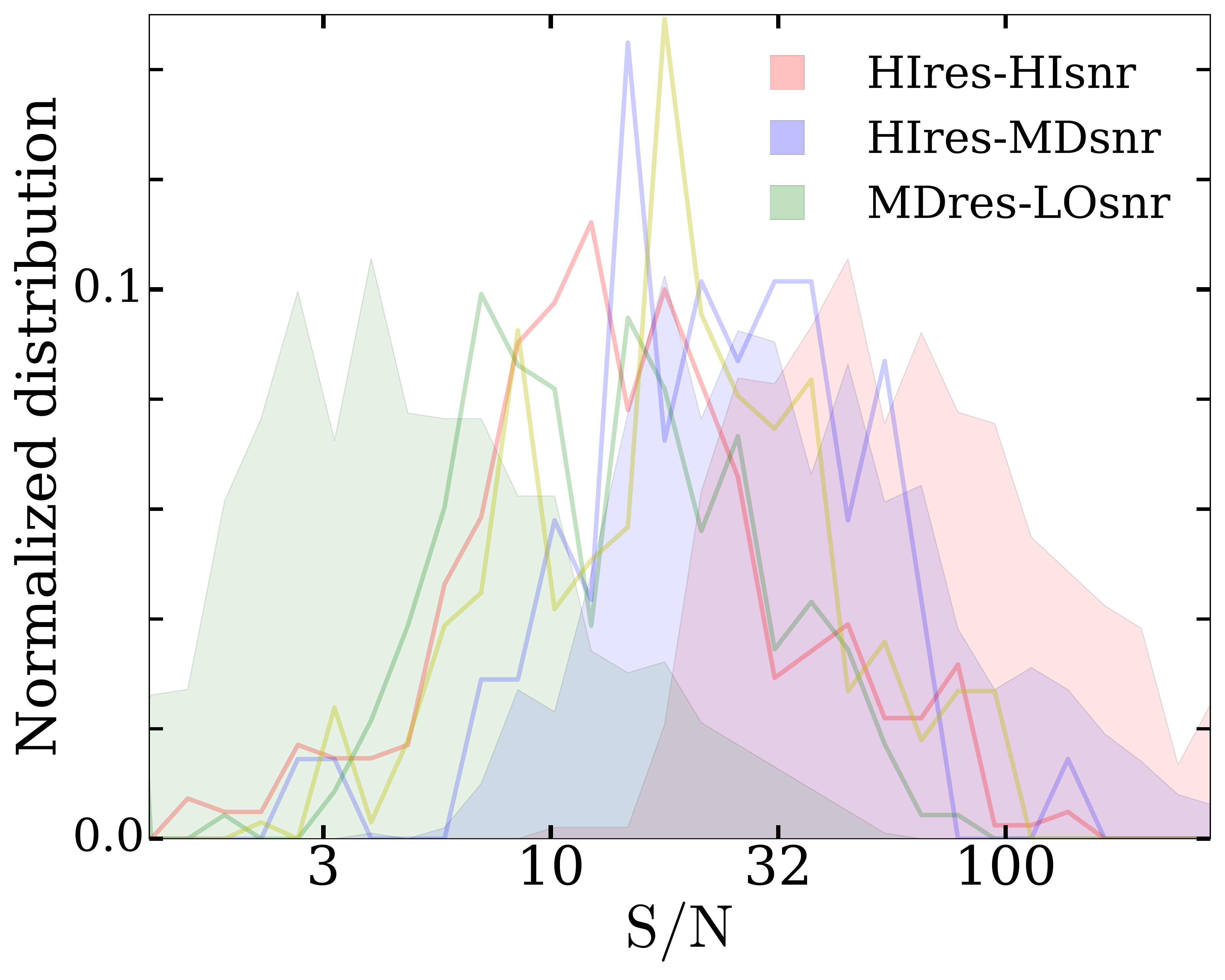}
\caption{The S/N distributions for the three mock quasar samples are denoted as {\it shaded regions}, and 
the distributions of the real samples are overplotted as {\it solid lines}, for comparison. The {\it blue region} 
broadly matches the HIRES sample distribution, and the {\it red} and {\it green areas} cover the tails of the 
distributions of the real samples.} 
\label{fig:mocksample}
\end{figure}

   For our purpose, we modify \texttt{QSOSim10} to create three sets of $1\,000$  
mock quasar spectra each, with different spectral resolution and S/N ratio broadly covering the range of values 
in our real data samples. We consider spectral resolutions $R=100\,000$ and $R=50\,000$, which we 
name ${\rm HI_{}res}$ and ${\rm MD_{}res}$, respectively, referring to high and medium resolution. Since 
\texttt{QSOSim10} computes BOSS noise for each spectrum, we adopt these BOSS values and simply apply a 
constant reduction factor for our mock samples, i.e., factors 3, 20 and 50, named ${\rm LO_{}snr}$, ${\rm 
MD_{}snr}$, and ${\rm HI_{}snr}$, respectively, accounting for low, medium and high S/N ratios. Table  
\ref{ta:mock} lists the median S/N ratio values, the resolution, and number of spectra for 
each mock sample. Figure \ref{fig:mocksample} illustrates the S/N ratio distributions as {\it shaded areas}, and  
overplotted as {\it solid lines} are those of the four real quasar samples. The mock 
${\rm HIres-MDsnr}$ sample ({\it blue shaded region}) broadly matches the HIRES sample ({\it blue line}), 
and the {\it shaded red} and {\it shaded green regions} cover the higher and lower tails, respectively, of the 
distributions of the real samples, which will be useful to determine the dependencies on S/N.

%-------------------------------------- Completeness and purity  ---------------------------------------
\subsubsection{Completeness and Purity in Mock Spectra Searches}\label{sec:mocksearch}

\begin{figure*}\center        % COMPLETENESS
\includegraphics[width=0.45\textwidth]{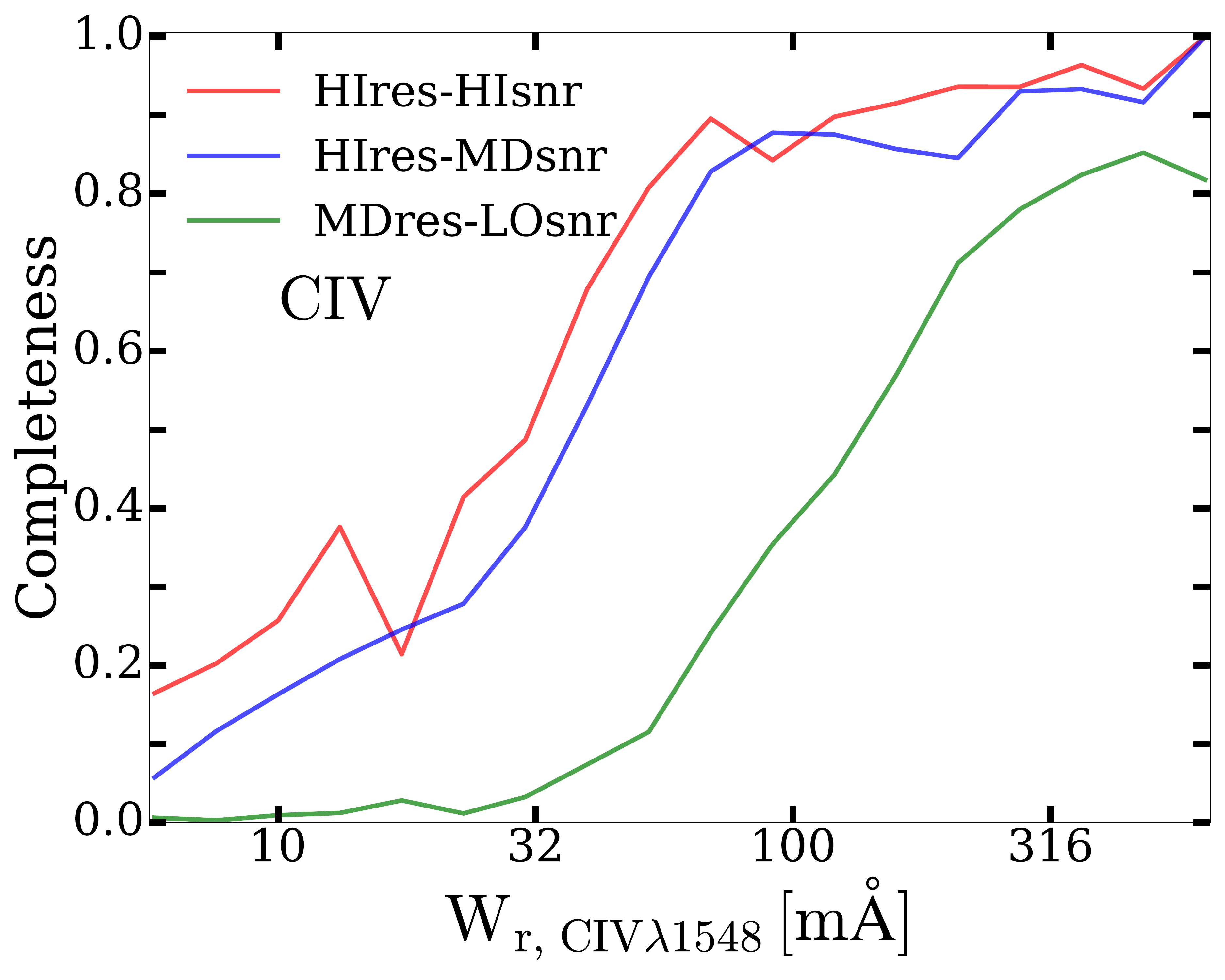}\includegraphics[width=0.45\textwidth]{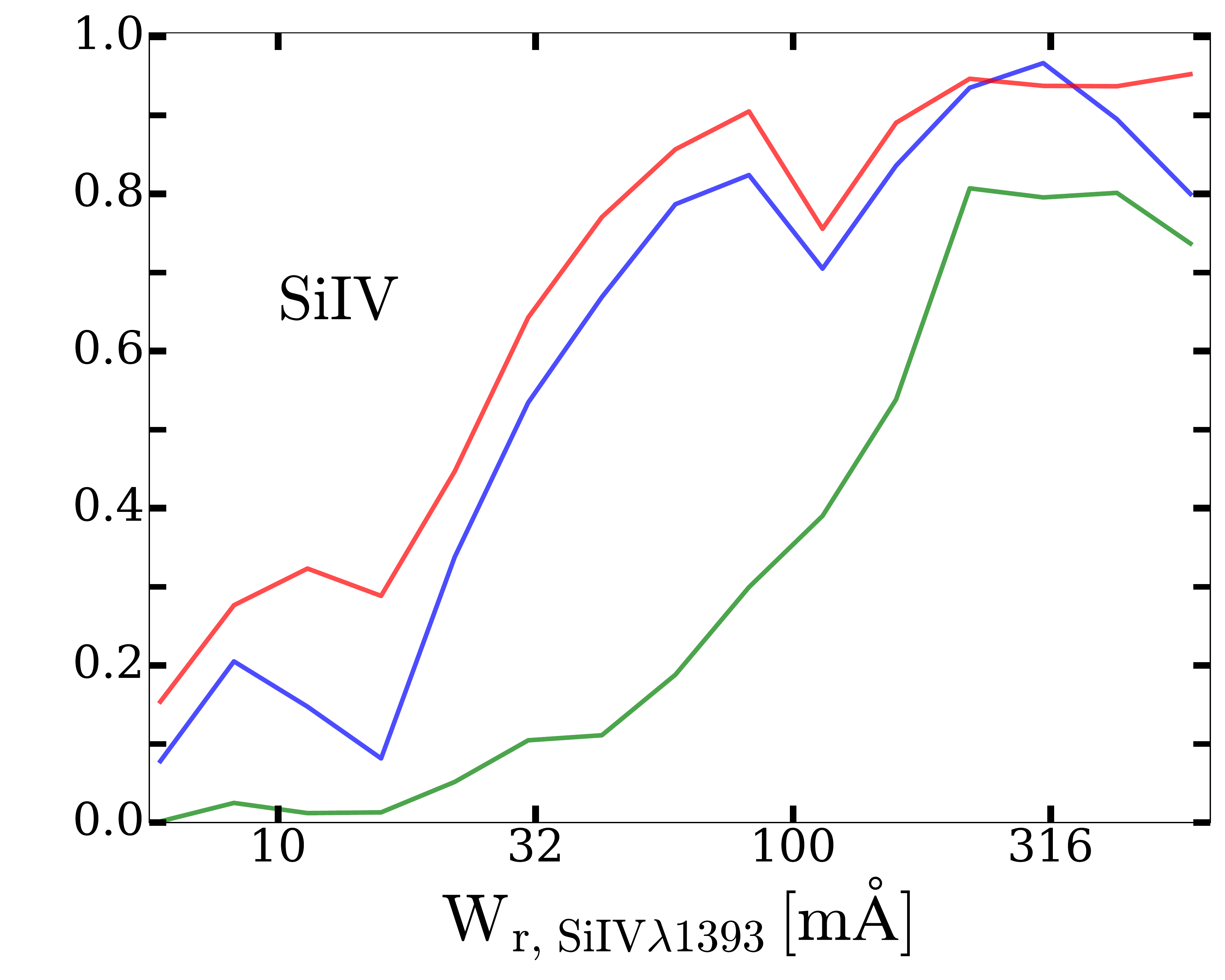}
\includegraphics[width=0.45\textwidth]{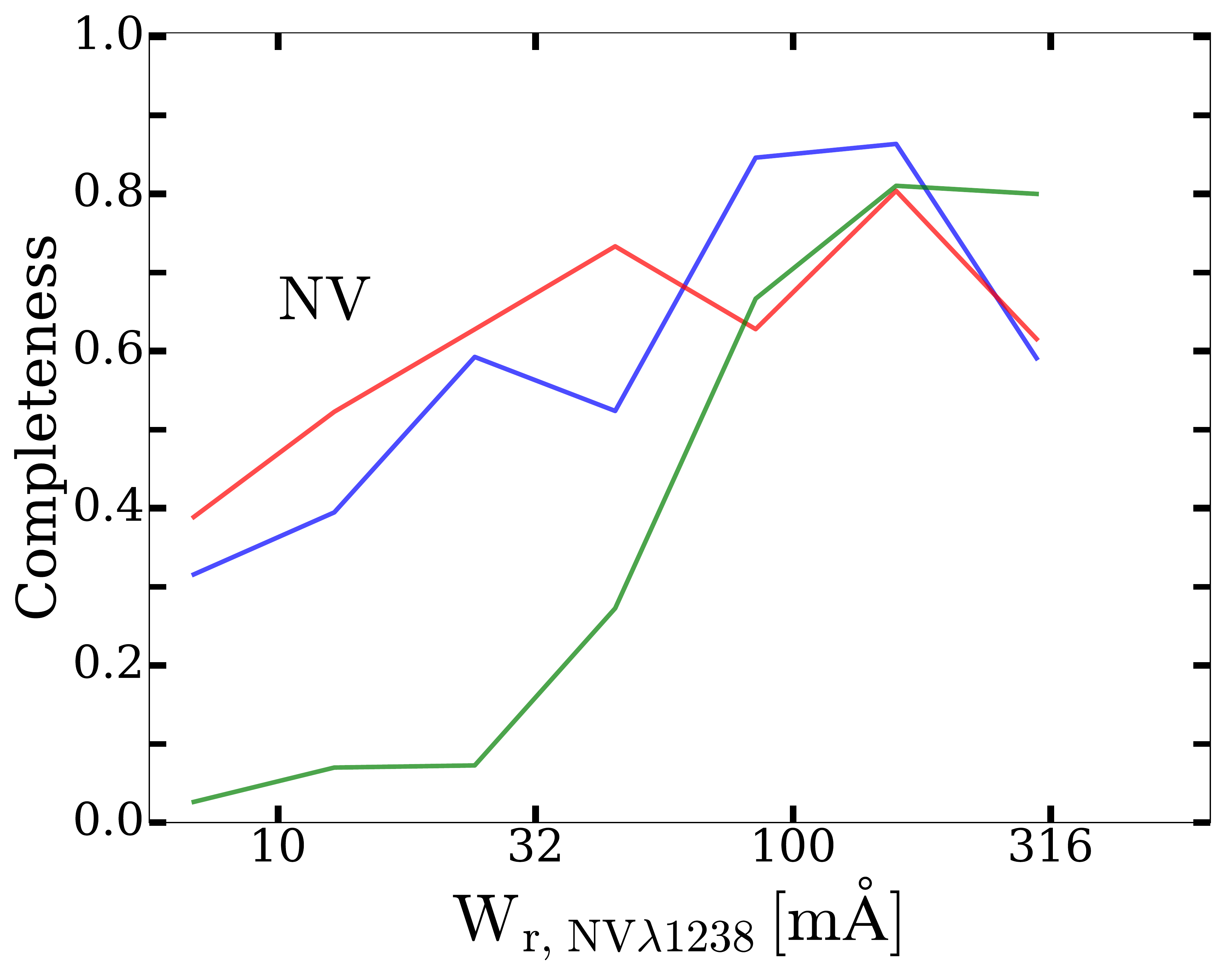}\includegraphics[width=0.45\textwidth]{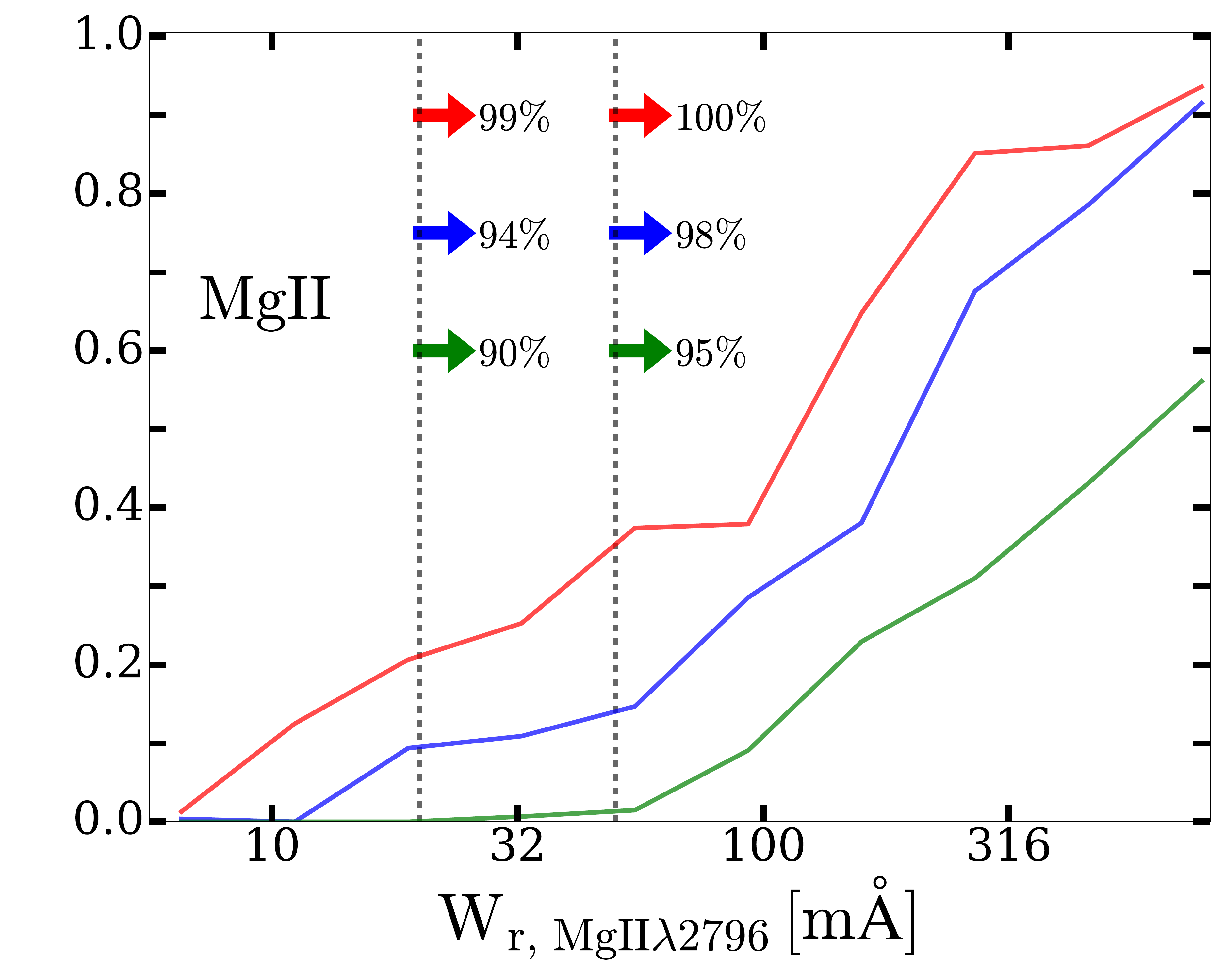}
\caption{Completeness fractions with respect to rest-frame equivalent width, for the low wavelength line of the 
CIV, SiIV, NV and MgII doublets, and for the three mock catalogs. The lower right panel also indicates the 
purity considering the absorbers with equivalent widths $W_r> 20\,{\rm m\angs}$ and $W_r> 50\,{\rm m\angs}$, 
for the three mock samples of MgII. The arrows are color coded as the lines. The horizontal axes are 
logarithmically spaced to facilitate the visualization.} 
\label{fig:completeness}
\end{figure*}

   We run our search code on the mock spectra to assess the purity and completeness of the results 
for all the species. We define the completeness as the fraction of mock absorbers, i.e., actual matches, 
that are detected by our code (considering only detections of at least 3$\sigma$ significance), and 
for several equivalent width bins. The purity denotes the ratio between the number of the doublets detected 
by our code that are real and the total number of detections. In all cases, we only consider the doublets 
outside the \lya forest. We have 
also used these mock spectra to explore and find the optimal search parameters.

   Figure \ref{fig:completeness} displays the completeness of the search as a function of the equivalent width 
of the bluer member of each doublet species, and for the three mock samples. The two highest S/N  
samples ({\it red} and {\it blue lines}) show little difference, with HIsnr 
sample ({\it red lines}) presenting a completeness $\sim 5-15\%$ higher than the MDsnr sample ({\it blue lines}), 
owing to the better S/N ratio of the spectra which allows for the detection of weaker features. For the lowest resolution 
and low S/N ratio sample ({\it green lines}), the completeness is, in general, significantly lower than that of the 
other two samples, especially at small equivalent width values. Given the distributions of S/N ratios for the real 
spectra, we expect most of these to show completeness curves between the green and red lines in Figure 
\ref{fig:completeness}.

%A maximum completeness of $\sim 60 - 80\%$ is generally 
%reached beyond $W_r\sim 50 - 100\,{\rm m\angs}$, depending in detail on the species and mock sample. Figure 
%\ref{fig:completeness} indicates a saturation of the completeness above $W_r\sim 100\,{\rm m\angs}$ for the 
%CIV, SiIV and NV species and, in some cases, a decrease is also observed at larger $W_r$ values. We have 
%tested that this effect is not connected to the searching procedure, but to the small number of absorption features 
%with an equivalent width above $W_r\sim 100\,{\rm m\angs}$ in the mock spectra (see the low fraction of 
%doublets above $W_r\sim 100\,{\rm m\angs}$ in the left panel of Figure \ref{fig:comcont}). The completeness 
%shows a rather positive slope, without saturation, for the MgII species up to $\sim 1\,{\rm \angs}\,$  because, by 
%construction, there are many MgII absorption lines with high equivalent widths. 

    The overall purity for each species and mock sample is indicated in Table 
\ref{ta:mock}, and is generally above $90\%$ for the CIV, SiIV and NV species, 
and $\gsim80\%$ for MgII. The last four columns show purity values considering 
only absorbers with equivalent width for the strongest 
line in the doublet  $W_r> 10\,{\rm m\angs}$, reaching values  
within $99-100\%$ in all CIV, SiIV and NV mock samples.  For MgII,  the lower 
right panel of Figure \ref{fig:completeness} indicates additional purity values for 
absorbers with equivalent widths $W_r> 20\,{\rm m\angs}$ and $W_r> 50\,{\rm 
m\angs}$, with the same color code as the lines in the plot. In general, purity 
values from the mock catalogs are higher than those computed from real spectra, 
likely because mocks have lower complexity, {\bf specially in the absorption line 
profiles and in the lack of effects, such as sky lines or outliers, that can contaminate 
pixels. } 

\begin{figure*}\center        %   CONTINUA
\includegraphics[width=0.98\textwidth]{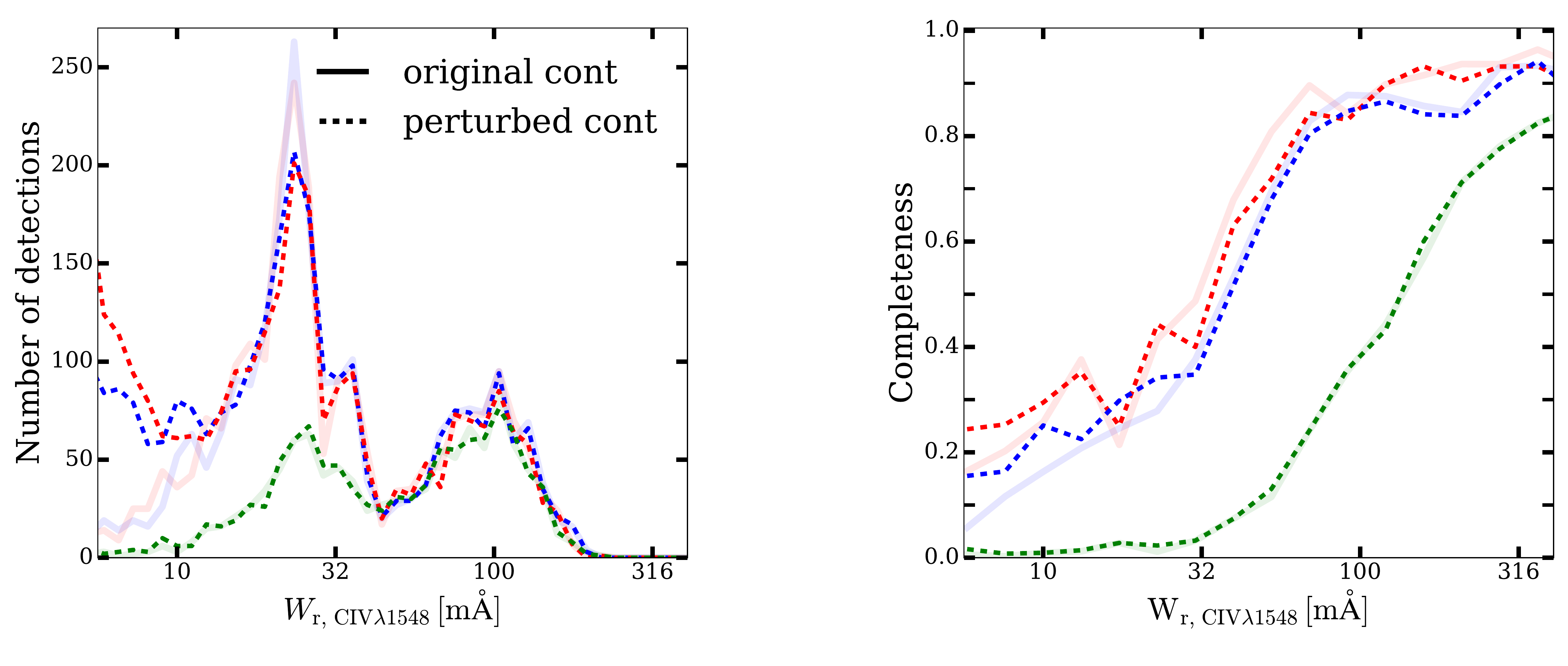}
\caption{ {\it Left panel:} Comparison between the number detections for the continuum-perturbed 
({\it dashed lines}) and 
unperturbed cases ({\it solid lines}). There is a slight  
increase in the number of detections at small equivalent widths ($W_r\lesssim 15\,{\rm m\angs}$) when the continuum variations are larger than 
the noise ({\it blue} and {\it red lines}). 
{\it Right panel:} Comparison of the completeness 
between the three mock quasar samples 
without and with variations in the quasar continua. The completeness 
increases slightly for the two high-resolution samples ({\it red} and {\it blue lines}) at low equivalent widths, although 
this trend is contributed by false positives misidentified 
as real doublets. The low S/N ratio sample ({\it green lines}) is not affected because of the small variations in the 
continuum.  The horizontal axes are logarithmically spaced to facilitate the visualization.}
\label{fig:comcont}
\end{figure*}

%%-------------------------------------- noise contamination  ---------------------------------------
%\subsubsection{Noise Contamination}\label{sec:noise}
%
%   We perform an additional test, running the search code on the mock spectra but 
%without metal absorption features, i.e., noise-only spectra, and compute the number of false-positive 
%doublet detections. This gives insight on the effect of noise to purity. We obtain only one false-positive 
%detection from the whole sample of mock spectra, a SiIV doublet in the noise-only MDres-LOsnr sample. 
%If the real noise is well reproduced by the mocks, we can safely assume that the fraction of 
%noise contamination in the real spectra is negligible. Most of the contamination must therefore come 
%from blends between lines and/or with noise, although real spectra may also contain other contaminants 
%not present in the mocks, e.g., cosmic rays, which can increase the false-positive detections compared 
%to the results from the test. We expect these effects to have little impact since our search code accounts 
%for the identification and correction of outliers in the spectra. 
%
%   We stress that the real quasar samples, HIRES, UVES, KODIAQ DR1 and KODIAQ DR2, are 
%constituted by quasars each with different values of spectral resolution and signal-to-noise ratio. 
%Therefore, the mock samples and their respective completeness and purity results do not directly   
%represent any of the real samples.

%-------------------------------------- CONTINUA TEST ---------------------------------------
\subsection{Continuum Placement Effects}\label{sec:testcont}

         Obtaining the quasar continuum is particularly difficult in the Lyman-alpha forest region of the spectra, where 
multiple hydrogen and metal absorption features overlap and reduce 
the average transmission. \cite{FaucherGiguere2008} estimated the continuum uncertainty in this region to be 
of a few percent in spectra with S/N$\sim 10$, increasing toward lower S/N and higher redshifts. Outside the 
Lyman-alpha forest, however, most of the spectrum is free of absorption and the continuum can be precisely 
calculated, its uncertainty being a small fraction of the noise in those regions. 

    To test the effect of continuum placement on the doublet search, we introduce variations in the continuum for 
each of our normalized mock spectra as 
follows: we consider eight evenly-spaced wavelength positions (nodes) in the normalized spectrum, the first 
and the last nodes corresponding to the first and last pixels, respectively. For every node, we then compute  
a small perturbation by randomly sampling a Gaussian distribution centered at 1, corresponding to 
the value of the nominal unabsorbed normalized flux, and standard deviation $\sigma_{\rm C}=0.01$, 
corresponding to a 1\% variation in the flux. Finally, the flux of the eight perturbed nodes is connected 
using linear interpolation, and the flux of each pixel in the spectrum is multiplied by the value of the 
interpolation at the corresponding pixel wavelength.  Selecting $\sigma_{\rm C}=0.01$ for the region 
outside the forest is likely an overestimation of the uncertainty in the continuum, especially for the high 
S/N spectra, but we adopt this value as an upper limit and interpret our results as a worst case. 

   Figure \ref{fig:comcont} shows the results after searching the $1\,000$ mock spectra of each sample 
with variations in the 
continua ({\it dashed lines}) and using the original unperturbed mocks ({\it solid lines}), for the case of 
CIV. The {\it left panel} shows the number of detections at every wavelength bin. 
The low S/N samples ({\it green lines}) show no differences between the perturbed and unperturbed cases 
because the noise dominates the signal and variations of the continuum are embedded within the noise. 
For the two high S/N cases, the overestimated continuum variations result in an increase of the number of low equivalent width doublets, 
$W_r\lesssim 15\,{\rm m\angs}$, detected in the search.  The small variations in the completeness shown 
in the {\it right panel} indicate that most of the newly detected features are not real. 

   Even assuming these  
large continuum variations, the results are unaffected at equivalent widths above 
$W_r\sim 15\,{\rm m\angs}$, concluding that the continuum uncertainties are not a concern for our analysis 
above this threshold.
%We have also explored the case using $\sigma_{\rm C}=0.005$, overall resulting in not 
%visible differences but with a decrease of the purity in the high-resolution cases down to $\sim44\%$ and 
%$\sim68\%$, respectively.
We have not observed apparent effects on the redshift distributions due to changes in the continua 
in any case.

%%--------------------------------------CATALOGS ---------------------------------------
\section{Metal Doublet Catalogs}\label{sec:catalogs}

    We present here the doublet catalogs resulting from our search. 
    
    As mentioned in \S~\ref{sec:data}, a fraction of the doublets in the samples are repeated because some 
spectra are included in various surveys, and, for  the case of the KODIAQ samples, the same quasar has  
several spectra from different observational campaigns. We consider  
two doublets to be the same when they accomplish all the following requirements: ({\it i}) the doublets 
appear in observations of the same quasar, ({\it ii}) the average redshifts of the doublets are offset by 
$\le 50\,{\rm km\,s^{-1}}$, implying redshift variations of $\Delta z/ (1+z) \sim 10^{-4}$, and ({\it iii}) the 
difference between the equivalent width of the respective absorption lines in the two doublets, for both lines 
of the doublet, is $\Delta W<20\%$. When two doublets are considered to be 
the same feature, we keep the one with the highest total detection significance, expressed as 
$W_{ l}/\sigma_{W_{ l}} + W_{ rg}/\sigma_{W_{ rg}}$, where $l$ and $rg$ denote the {\bf lower} and {\bf higher 
wavelength} lines in the doublet, respectively. Variations of a factor of two in these two threshold values lead to  
variations of a few percent in the total number of doublets.

\begin{table}\center
	\begin{center}
	\caption{Doublet Catalogs}\label{ta:cats}
	\begin{threeparttable}
		\begin{tabular}{ccccc} 
		\hline
		\hline
		                     & \multicolumn{4}{c}{Number of doublets }             \\
		Sample				&CIV       		&SiIV            &NV    &MgII       \\ 
		\hline 
		UVES                        &$3\,291$		&$1\,394$			&$156$	&$6\,482$					\\
		HIRES			&$849$		&$290$			&$25$	&$481$ 						    \\
		KODIAQ DR1		&$1\,190$		&$430$			&$34$	&$550$						\\
		KODIAQ DR2		&$1\,515$		&$598$			&$75$	&$1\,004$						\\
		\hline  
		Total				&$5\,656$		&$2\,258$			&$239$	&$7\,919$						\\	
		\hline
		\end{tabular}		
	\end{threeparttable}
	\end{center}
\end{table}

\begin{figure*}\center      % W + Z DISTR  TOTAL 
\includegraphics[width=1\textwidth]{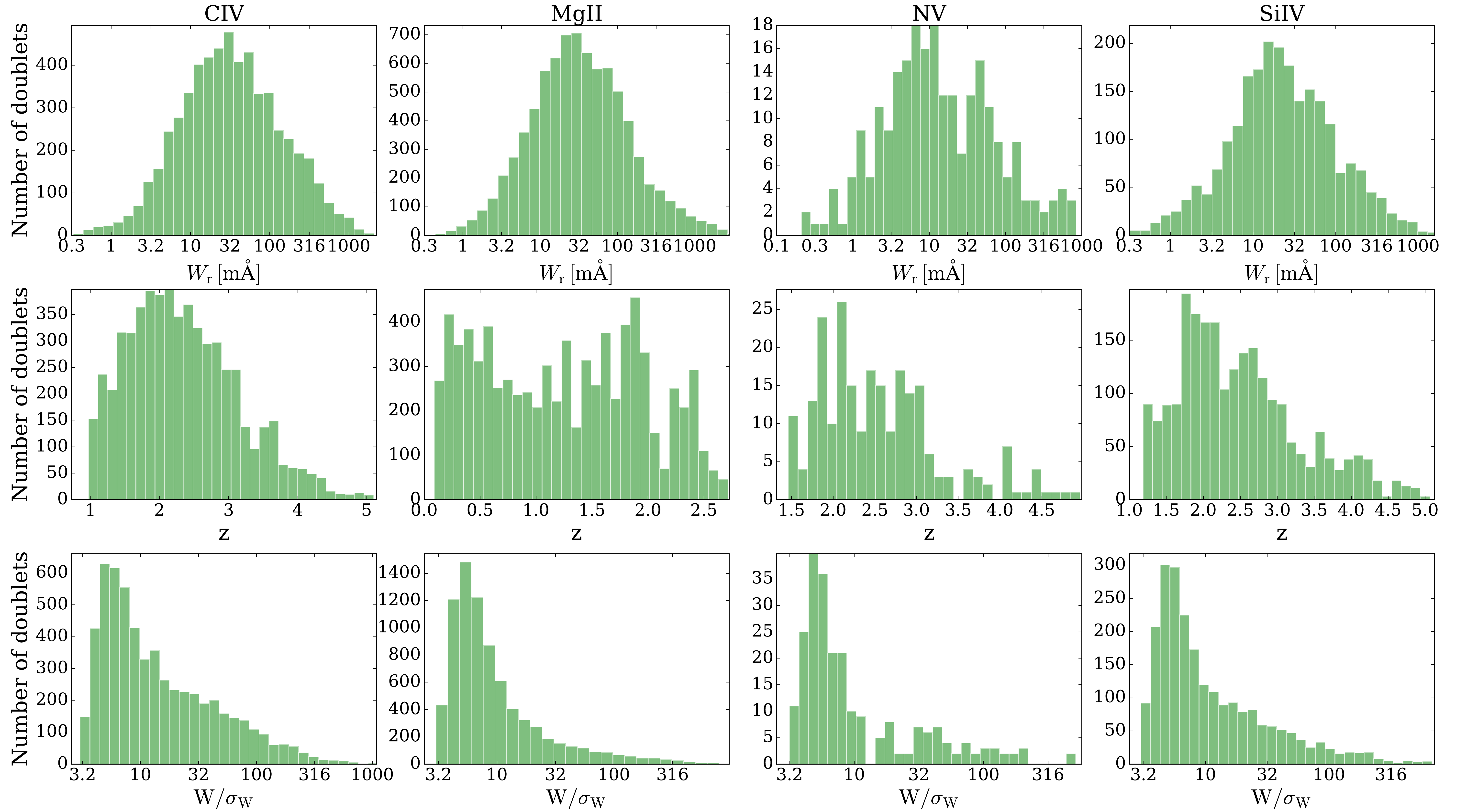}
\caption{Rest equivalent width, $W_r$, redshift, $z$, and detection significance, $W/\sigma_W$, 
distributions for the {\bf higher frequency} line of the doublets in the total catalogs. The horizontal axes 
for equivalent width and detection significance are logarithmically spaced to facilitate the visualization.} 
\label{fig:tot_distr}
\end{figure*}

   Figure \ref{fig:tot_distr} shows the distributions of rest-frame equivalent width, $W_r$, doublet 
redshift, $z$, and significance of the detection, $W/\sigma_W$, for the {\bf lower wavelength} lines in the 
doublet, and the four species in the final total catalog, including the spectra from all samples. 
The redshift of the doublets is computed as the weighted 
mean of the redshift of the two lines in the doublet, using the significance of the detections as the 
weighting parameter. Table \ref{ta:cats} quotes the number of doublets for each species and for each catalog, 
and in the final catalog after accounting for the repeated doublets.   

   Table \ref{ta:civ} illustrates the structure of the final CIV catalog available at \url{https://github.com/lluism/OMG}. 
The name, right ascension and declination of the quasar are quoted in the first to the third 
columns, respectively, and the average redshift of the doublet is displayed in the fourth column. The 
fifth to the tenth columns denote the rest-frame equivalent width and  its uncertainty (in ${\rm 
m\angs}$), and detection significance, for the CIV$\,\lambda$1548  and CIV$\,\lambda$1550 lines, 
respectively. We list the detection significances computed from the wavelength and uncertainty values in 
Angstrom, at the observer frame using the redshift for each individual line, and before rounding. In some 
cases, these values differ from the simple division of the values quoted in the table. 
The last column denotes the quasar sample from where the doublet is obtained. The 
individual catalogs and distribution plots for each species and quasar sample are publicly 
available at the link indicated above.

%%--------------------------------------CONCLUSION ---------------------------------------
\section{Conclusions}\label{sec:conclusion}

     In this first paper of the series Origin of Metals around Galaxies (OMG), we build and 
publicly release the large metal-line doublet catalogs that we will analyse in 
future papers. 

\begin{table*}\center
	\begin{center}
	\caption{Example CIV doublet catalog\label{ta:civ}}
	\begin{threeparttable}
		\begin{tabular}{ccccccccccc} 
		\hline
		\hline
		Quasar 		   	     &RA        	    &Dec    &$\bar z$ 	&$W_{r,1548}$	&$\sigma_{W_{r,1548}}$  	&$W_{48}/\sigma_{W_{48}}$          &$W_{r,1550}$	&$\sigma_{W_{r,1550}}$  	&$W_{50}/\sigma_{W_{50}}$                &Sample\\ \hline
J082540+354414  &08:25:40.12  &+35:44:14.20   &$3.073$   &$7.81$    &$1.21$   &$6.22$       &$5.31$  &$1.11$  &$4.95$    &KODIAQ  \\ 
J000150-015940  &00:01:50.00  &-01:59:40.00   &$1.023$   &$99.52$    &$20.40$   &$4.87$     &$66.32$  &$17.31$  &$3.82$   &UVES  \\ 
J045142-132033  &04:51:42.60  &-13:20:33.00   &$2.847$   &$27.41$    &$1.40$   &$19.55$      &$11.60$  &$1.33$  &$8.95$    &HIRES  \\ 

...  &  &   &   &    &   &    &   &  &    &  \\ 
		\tableline	
		\end{tabular}
	\end{threeparttable}
	\tablecomments{
	First column: Quasar name. 
	Second column: Right ascension.
        Third column: Declination. 
	Fourth column: average redshift of the doublet.
	Fifth column: Rest-frame equivalent width of CIV$\,\lambda$1548 (in ${\rm m\angs}$).
        Sixth column: Rest-frame uncertainty for the equivalent width of CIV$\,\lambda$1548 (in ${\rm m\angs}$).
        Seventh column: Detection significance  for the CIV$\,\lambda$1548 line.
	Eigth column: Rest-frame equivalent width of CIV$\,\lambda$1550 (in ${\rm m\angs}$).
        Ninth column: Rest-frame uncertainty for the equivalent width of CIV$\,\lambda$1550 (in ${\rm m\angs}$).
        Tenth column: Detection significance  for the CIV$\,\lambda$1550 line.
        Eleventh column: Quasar sample from where the doublet is obtained}	
	\end{center}
\end{table*}

    We have designed a metal-doublet search algorithm, based on that by \cite{Cooksey2008}, that 
searches for metal-line doublets of CIV$\,\lambda\lambda$1548,1550, MgII$\,\lambda\lambda$2796,
2803, SiIV$\,\lambda\lambda$1393,1402, and NV$\,\lambda\lambda$1238,1242, in quasar spectra  
in a fully-automatic fashion, i.e., without human intervention. We have tested the purity of the results 
by looking for false doublets using line separations larger than the true ones in real spectra, and have  
created three large samples of mock quasar spectra with different resolution and S/N ratio values 
to estimate the completeness.  We have also assessed the impact of variations in the determination of the intrinsic 
quasar continua on the purity.  Finally, we have applied our code to the rest-frame wavelength ranges outside 
the Lyman-alpha forest of  the spectra of 
690 quasars, compiled from the KODIAQ datasets and observations with the UVES and HIRES 
instruments. Our findings can be summarized as follows:

\begin{itemize}

\item We build catalogs with $5\,656$ CIV$\,\lambda\lambda$1548,
	1550 doublets, $7\,919$ doublets of MgII$\,\lambda\lambda$2796,2803, 
	$2\,258$ of SiIV$\,\lambda\lambda$1393,1402, and 239 of NV$\,\lambda\lambda$1238,1242, 
	available at \url{https://github.com/lluism/OMG}.

\item We present the purity of the individual samples with equivalent width, redshift, and significance 
         of the detection, the latter being the parameter that purity depends on the most. 
	
\item  We expect contamination to the purity arising from errors in the calculation of the continuum to be 
         negligible outside the Lyman-alpha forest. In cases of large continuum errors, if any, this effect may 
         introduce false positive detections with small equivalent widths ($\lesssim 15\,{\rm m\angs}$) in 
         spectra with high signal-to-noise ratios (${\rm S/N}\gtrsim10$). 
        
\end{itemize}

   In the upcoming work, we will perform the cross-correlation of the weak CIV absorbers found here 
with the \lya forest from BOSS. These calculations will enable us to obtain the bias factor of the 
intergalactic absorbers at different equivalent widths, and thus place constraints on their progenitor 
galaxy population.

%-------------------------------------- ACKNOWLEDGEMENTS ---------------------------------------
\section*{Acknowledgements}

We thank the referee for useful comments that improved the clarity of our work. The authors are grateful to 
Kathy Cooksey for running her code on our test spectrum, 
providing and allowing us to use her code, and for useful answers to our many questions.  We also wish 
to thank H\'ector Hiss, David Bowen, Valentina D' Odorico and Matt Pieri for their comments and suggestions. 
We acknowledge Wallace Sargent, Tom Barlow, Michael Rauch, and Robert Simcoe for sharing their Keck 
spectra, and thank the following for their work on spectral reduction and normalisation: A. Chaudhary, A. Gormley, 
A. Hartstone, B. Zych , D. Dougan, D. Mountford, G. Kacprzak, H. Campbell , H. Menager, J. Diacoumis, 
J. Ouellet, J .Sinclair, K. Ali, M. Bainbridge,  N. Harvey, R. Buning, R. Carswell, R. McSwiney, T. Wei, and 
V. Efremova.

%-----------------------BIBLIOGRAPHY ----------------------------
\newpage
\bibliographystyle{apj}
\bibliography{catalog}\label{References}

\begin{thebibliography}{}
\expandafter\ifx\csname natexlab\endcsname\relax\def\natexlab#1{#1}\fi

\bibitem[{{Adelberger} {et~al.}(2005){Adelberger}, {Shapley}, {Steidel},
  {Pettini}, {Erb}, \& {Reddy}}]{Adelberger2005}
{Adelberger}, K.~L., {Shapley}, A.~E., {Steidel}, C.~C., {et~al.} 2005, \apj,
  629, 636

\bibitem[{{Adelberger} {et~al.}(2003){Adelberger}, {Steidel}, {Shapley}, \&
  {Pettini}}]{Adelberger2003}
{Adelberger}, K.~L., {Steidel}, C.~C., {Shapley}, A.~E., \& {Pettini}, M. 2003,
  \apj, 584, 45

\bibitem[{{Alam} {et~al.}(2015){Alam}, {Albareti}, {Allende Prieto}, {Anders},
  {Anderson}, {Anderton}, {Andrews}, {Armengaud}, {Aubourg}, {Bailey}, \&
  et~al.}]{Alam2015}
{Alam}, S., {Albareti}, F.~D., {Allende Prieto}, C., {et~al.} 2015, \apjs, 219,
  12

\bibitem[{{Bagdonaite} {et~al.}(2014){Bagdonaite}, {Ubachs}, {Murphy}, \&
  {Whitmore}}]{Bagdonaite2014}
{Bagdonaite}, J., {Ubachs}, W., {Murphy}, M.~T., \& {Whitmore}, J.~B. 2014,
  \apj, 782, 10

\bibitem[{{Bautista} {et~al.}(2017){Bautista}, {Busca}, {Guy}, {Rich},
  {Blomqvist}, {du Mas des Bourboux}, {Pieri}, {Font-Ribera}, {Bailey},
  {Delubac}, {Kirkby}, {Le Goff}, {Margala}, {Slosar}, {Vazquez}, {Brownstein},
  {Dawson}, {Eisenstein}, {Miralda-Escud{\'e}}, {Noterdaeme},
  {Palanque-Delabrouille}, {P{\^a}ris}, {Petitjean}, {Ross}, {Schneider},
  {Weinberg}, \& {Y{\`e}che}}]{Bautista2017}
{Bautista}, J.~E., {Busca}, N.~G., {Guy}, J., {et~al.} 2017, \aap, 603, A12

\bibitem[{{Bertone} {et~al.}(2005){Bertone}, {Stoehr}, \&
  {White}}]{Bertone2005}
{Bertone}, S., {Stoehr}, F., \& {White}, S.~D.~M. 2005, \mnras, 359, 1201

\bibitem[{{Blomqvist} {et~al.}(2015){Blomqvist}, {Kirkby}, {Bautista},
  {Arinyo-i-Prats}, {Busca}, {Miralda-Escud{\'e}}, {Slosar}, {Font-Ribera},
  {Margala}, {Schneider}, \& {Vazquez}}]{Blomqvist2015}
{Blomqvist}, M., {Kirkby}, D., {Bautista}, J.~E., {et~al.} 2015, \jcap, 11, 034

\bibitem[{Bouch{\'e} {et~al.}(2007)Bouch{\'e}, Lehnert, Aguirre, Péroux, \&
  Bergeron}]{Bouche2007}
Bouch{\'e}, N., Lehnert, M.~D., Aguirre, A., Péroux, C., \& Bergeron, J. 2007,
  Monthly Notices of the Royal Astronomical Society, 378, 525

\bibitem[{{Bouch{\'e}} {et~al.}(2006){Bouch{\'e}}, {Lehnert}, \&
  {P{\'e}roux}}]{Bouche2006}
{Bouch{\'e}}, N., {Lehnert}, M.~D., \& {P{\'e}roux}, C. 2006, \mnras, 367, L16

\bibitem[{{Busca} {et~al.}(2013){Busca}, {Delubac}, {Rich}, {Bailey},
  {Font-Ribera}, {Kirkby}, {Le Goff}, {Pieri}, {Slosar}, {Aubourg}, {Bautista},
  {Bizyaev}, {Blomqvist}, {Bolton}, {Bovy}, {Brewington}, {Borde}, {Brinkmann},
  {Carithers}, {Croft}, {Dawson}, {Ebelke}, {Eisenstein}, {Hamilton}, {Ho},
  {Hogg}, {Honscheid}, {Lee}, {Lundgren}, {Malanushenko}, {Malanushenko},
  {Margala}, {Maraston}, {Mehta}, {Miralda-Escud{\'e}}, {Myers}, {Nichol},
  {Noterdaeme}, {Olmstead}, {Oravetz}, {Palanque-Delabrouille}, {Pan},
  {P{\^a}ris}, {Percival}, {Petitjean}, {Roe}, {Rollinde}, {Ross}, {Rossi},
  {Schlegel}, {Schneider}, {Shelden}, {Sheldon}, {Simmons}, {Snedden},
  {Tinker}, {Viel}, {Weaver}, {Weinberg}, {White}, {Y{\`e}che}, \&
  {York}}]{Busca2013}
{Busca}, N.~G., {Delubac}, T., {Rich}, J., {et~al.} 2013, \aap, 552, A96

\bibitem[{{Cooksey} {et~al.}(2013){Cooksey}, {Kao}, {Simcoe}, {O'Meara}, \&
  {Prochaska}}]{Cooksey2013}
{Cooksey}, K.~L., {Kao}, M.~M., {Simcoe}, R.~A., {O'Meara}, J.~M., \&
  {Prochaska}, J.~X. 2013, \apj, 763, 37

\bibitem[{{Cooksey} {et~al.}(2008){Cooksey}, {Prochaska}, {Chen}, {Mulchaey},
  \& {Weiner}}]{Cooksey2008}
{Cooksey}, K.~L., {Prochaska}, J.~X., {Chen}, H.-W., {Mulchaey}, J.~S., \&
  {Weiner}, B.~J. 2008, \apj, 676, 262

\bibitem[{{Cooksey} {et~al.}(2011){Cooksey}, {Prochaska}, {Thom}, \&
  {Chen}}]{Cooksey2011}
{Cooksey}, K.~L., {Prochaska}, J.~X., {Thom}, C., \& {Chen}, H.-W. 2011, \apj,
  729, 87

\bibitem[{{Cooksey} {et~al.}(2010){Cooksey}, {Thom}, {Prochaska}, \&
  {Chen}}]{Cooksey2010}
{Cooksey}, K.~L., {Thom}, C., {Prochaska}, J.~X., \& {Chen}, H.-W. 2010, \apj,
  708, 868

\bibitem[{{Dawson} {et~al.}(2013){Dawson}, {Schlegel}, {Ahn}, {Anderson},
  {Aubourg}, {Bailey}, {Barkhouser}, {Bautista}, {Beifiori}, {Berlind},
  {Bhardwaj}, {Bizyaev}, {Blake}, {Blanton}, {Blomqvist}, {Bolton}, {Borde},
  {Bovy}, {Brandt}, {Brewington}, {Brinkmann}, {Brown}, {Brownstein}, {Bundy},
  {Busca}, {Carithers}, {Carnero}, {Carr}, {Chen}, {Comparat}, {Connolly},
  {Cope}, {Croft}, {Cuesta}, {da Costa}, {Davenport}, {Delubac}, {de Putter},
  {Dhital}, {Ealet}, {Ebelke}, {Eisenstein}, {Escoffier}, {Fan}, {Filiz Ak},
  {Finley}, {Font-Ribera}, {G{\'e}nova-Santos}, {Gunn}, {Guo}, {Haggard},
  {Hall}, {Hamilton}, {Harris}, {Harris}, {Ho}, {Hogg}, {Holder}, {Honscheid},
  {Huehnerhoff}, {Jordan}, {Jordan}, {Kauffmann}, {Kazin}, {Kirkby}, {Klaene},
  {Kneib}, {Le Goff}, {Lee}, {Long}, {Loomis}, {Lundgren}, {Lupton}, {Maia},
  {Makler}, {Malanushenko}, {Malanushenko}, {Mandelbaum}, {Manera}, {Maraston},
  {Margala}, {Masters}, {McBride}, {McDonald}, {McGreer}, {McMahon}, {Mena},
  {Miralda-Escud{\'e}}, {Montero-Dorta}, {Montesano}, {Muna}, {Myers},
  {Naugle}, {Nichol}, {Noterdaeme}, {Nuza}, {Olmstead}, {Oravetz}, {Oravetz},
  {Owen}, {Padmanabhan}, {Palanque-Delabrouille}, {Pan}, {Parejko},
  {P{\^a}ris}, {Percival}, {P{\'e}rez-Fournon}, {P{\'e}rez-R{\`a}fols},
  {Petitjean}, {Pfaffenberger}, {Pforr}, {Pieri}, {Prada}, {Price-Whelan},
  {Raddick}, {Rebolo}, {Rich}, {Richards}, {Rockosi}, {Roe}, {Ross}, {Ross},
  {Rossi}, {Rubi{\~n}o-Martin}, {Samushia}, {S{\'a}nchez}, {Sayres}, {Schmidt},
  {Schneider}, {Sc{\'o}ccola}, {Seo}, {Shelden}, {Sheldon}, {Shen}, {Shu},
  {Slosar}, {Smee}, {Snedden}, {Stauffer}, {Steele}, {Strauss}, {Streblyanska},
  {Suzuki}, {Swanson}, {Tal}, {Tanaka}, {Thomas}, {Tinker}, {Tojeiro},
  {Tremonti}, {Vargas Maga{\~n}a}, {Verde}, {Viel}, {Wake}, {Watson}, {Weaver},
  {Weinberg}, {Weiner}, {West}, {White}, {Wood-Vasey}, {Yeche}, {Zehavi},
  {Zhao}, \& {Zheng}}]{Dawson2013}
{Dawson}, K.~S., {Schlegel}, D.~J., {Ahn}, C.~P., {et~al.} 2013, \aj, 145, 10

\bibitem[{{Dekker} {et~al.}(2000){Dekker}, {D'Odorico}, {Kaufer}, {Delabre}, \&
  {Kotzlowski}}]{Dekker2000}
{Dekker}, H., {D'Odorico}, S., {Kaufer}, A., {Delabre}, B., \& {Kotzlowski}, H.
  2000, in \procspie, Vol. 4008, Optical and IR Telescope Instrumentation and
  Detectors, ed. M.~{Iye} \& A.~F. {Moorwood}, 534--545

\bibitem[{{Delubac} {et~al.}(2015){Delubac}, {Bautista}, {Busca}, {Rich},
  {Kirkby}, {Bailey}, {Font-Ribera}, {Slosar}, {Lee}, {Pieri}, {Hamilton},
  {Aubourg}, {Blomqvist}, {Bovy}, {Brinkmann}, {Carithers}, {Dawson},
  {Eisenstein}, {Gontcho}, {Kneib}, {Le Goff}, {Margala}, {Miralda-Escud{\'e}},
  {Myers}, {Nichol}, {Noterdaeme}, {O'Connell}, {Olmstead},
  {Palanque-Delabrouille}, {P{\^a}ris}, {Petitjean}, {Ross}, {Rossi},
  {Schlegel}, {Schneider}, {Weinberg}, {Y{\`e}che}, \& {York}}]{Delubac2015}
{Delubac}, T., {Bautista}, J.~E., {Busca}, N.~G., {et~al.} 2015, \aap, 574, A59

\bibitem[{{Eisenstein} {et~al.}(2011){Eisenstein}, {Weinberg}, {Agol},
  {Aihara}, {Allende Prieto}, {Anderson}, {Arns}, {Aubourg}, {Bailey},
  {Balbinot}, \& et~al.}]{Eisenstein2011}
{Eisenstein}, D.~J., {Weinberg}, D.~H., {Agol}, E., {et~al.} 2011, \aj, 142, 72

\bibitem[{{Faucher-Gigu{\`e}re} {et~al.}(2008){Faucher-Gigu{\`e}re}, {Lidz},
  {Hernquist}, \& {Zaldarriaga}}]{FaucherGiguere2008}
{Faucher-Gigu{\`e}re}, C.-A., {Lidz}, A., {Hernquist}, L., \& {Zaldarriaga}, M.
  2008, \apj, 688, 85

\bibitem[{{Ferland} {et~al.}(2013){Ferland}, {Porter}, {van Hoof}, {Williams},
  {Abel}, {Lykins}, {Shaw}, {Henney}, \& {Stancil}}]{Ferland2013}
{Ferland}, G.~J., {Porter}, R.~L., {van Hoof}, P.~A.~M., {et~al.} 2013, \rmxaa,
  49, 137

\bibitem[{{Font-Ribera} {et~al.}(2012){Font-Ribera}, {Miralda-Escud{\'e}},
  {Arnau}, {Carithers}, {Lee}, {Noterdaeme}, {P{\^a}ris}, {Petitjean}, {Rich},
  {Rollinde}, {Ross}, {Schneider}, {White}, \& {York}}]{Fontribera2012}
{Font-Ribera}, A., {Miralda-Escud{\'e}}, J., {Arnau}, E., {et~al.} 2012, \jcap,
  11, 059

\bibitem[{{Fox} {et~al.}(2005){Fox}, {Savage}, \& {Wakker}}]{Fox2005}
{Fox}, A.~J., {Savage}, B.~D., \& {Wakker}, B.~P. 2005, \aj, 130, 2418

\bibitem[{{Gontcho} {et~al.}(2017){Gontcho}, {Miralda-Escud{\'e}},
  {Font-Ribera}, {Blomqvist}, {Busca}, \& {Rich}}]{Gontcho2017}
{Gontcho}, S.~G.~A., {Miralda-Escud{\'e}}, J., {Font-Ribera}, A., {et~al.}
  2017, ArXiv e-prints, arXiv:1712.09886

\bibitem[{{Hayward} \& {Hopkins}(2017)}]{Hayward2017}
{Hayward}, C.~C., \& {Hopkins}, P.~F. 2017, \mnras, 465, 1682

\bibitem[{{Kacprzak} {et~al.}(2011){Kacprzak}, {Churchill}, {Barton}, \&
  {Cooke}}]{Kacprzak2011}
{Kacprzak}, G.~G., {Churchill}, C.~W., {Barton}, E.~J., \& {Cooke}, J. 2011,
  \apj, 733, 105

\bibitem[{{King} {et~al.}(2012){King}, {Webb}, {Murphy}, {Flambaum},
  {Carswell}, {Bainbridge}, {Wilczynska}, \& {Koch}}]{King2012}
{King}, J.~A., {Webb}, J.~K., {Murphy}, M.~T., {et~al.} 2012, \mnras, 422, 3370

\bibitem[{{Martin} {et~al.}(2010){Martin}, {Scannapieco}, {Ellison}, {Hennawi},
  {Djorgovski}, \& {Fournier}}]{Martin2010}
{Martin}, C.~L., {Scannapieco}, E., {Ellison}, S.~L., {et~al.} 2010, \apj, 721,
  174

\bibitem[{{Mas-Ribas} {et~al.}(2017){Mas-Ribas}, {Miralda-Escud{\'e}},
  {P{\'e}rez-R{\`a}fols}, {Arinyo-i-Prats}, {Noterdaeme}, {Petitjean},
  {Schneider}, {York}, \& {Ge}}]{Masribas2016c}
{Mas-Ribas}, L., {Miralda-Escud{\'e}}, J., {P{\'e}rez-R{\`a}fols}, I., {et~al.}
  2017, \apj, 846, 4

\bibitem[{{Mathes} {et~al.}(2017){Mathes}, {Churchill}, \&
  {Murphy}}]{Mathes2017}
{Mathes}, N.~L., {Churchill}, C.~W., \& {Murphy}, M.~T. 2017, ArXiv e-prints,
  arXiv:1701.05624

\bibitem[{Milakovic {et~al.}(2017)Milakovic, Webb, \& Dumont}]{Qsosim10}
Milakovic, D., Webb, J.~K., \& Dumont, V. 2017, QSOSIM 10: Simulated Quasar
  Spectra Generator, doi:10.5281/zenodo.439394

\bibitem[{{Murphy} {et~al.}(2003){Murphy}, {Webb}, \& {Flambaum}}]{Murphy2003}
{Murphy}, M.~T., {Webb}, J.~K., \& {Flambaum}, V.~V. 2003, \mnras, 345, 609

\bibitem[{{O'Meara} {et~al.}(2017){O'Meara}, {Lehner}, {Howk}, {Prochaska},
  {Fox}, {Peeples}, {Tumlinson}, \& {O'Shea}}]{Omeara2017}
{O'Meara}, J.~M., {Lehner}, N., {Howk}, J.~C., {et~al.} 2017, ArXiv e-prints,
  arXiv:1707.07905

\bibitem[{{O'Meara} {et~al.}(2015){O'Meara}, {Lehner}, {Howk}, {Prochaska},
  {Fox}, {Swain}, {Gelino}, {Berriman}, \& {Tran}}]{Omeara2015}
---. 2015, \aj, 150, 111

\bibitem[{{Oppenheimer} {et~al.}(2012){Oppenheimer}, {Dav{\'e}}, {Katz},
  {Kollmeier}, \& {Weinberg}}]{Oppenheimer2012}
{Oppenheimer}, B.~D., {Dav{\'e}}, R., {Katz}, N., {Kollmeier}, J.~A., \&
  {Weinberg}, D.~H. 2012, \mnras, 420, 829

\bibitem[{{P{\^a}ris} {et~al.}(2014){P{\^a}ris}, {Petitjean}, {Aubourg},
  {Ross}, {Myers}, {Streblyanska}, {Bailey}, {Hall}, {Strauss}, {Anderson},
  {Bizyaev}, {Borde}, {Brinkmann}, {Bovy}, {Brandt}, {Brewington},
  {Brownstein}, {Cook}, {Ebelke}, {Fan}, {Filiz Ak}, {Finley}, {Font-Ribera},
  {Ge}, {Hamann}, {Ho}, {Jiang}, {Kinemuchi}, {Malanushenko}, {Malanushenko},
  {Marchante}, {McGreer}, {McMahon}, {Miralda-Escud{\'e}}, {Muna},
  {Noterdaeme}, {Oravetz}, {Palanque-Delabrouille}, {Pan}, {Perez-Fournon},
  {Pieri}, {Riffel}, {Schlegel}, {Schneider}, {Simmons}, {Viel}, {Weaver},
  {Wood-Vasey}, {Y{\`e}che}, \& {York}}]{Paris2014}
{P{\^a}ris}, I., {Petitjean}, P., {Aubourg}, {\'E}., {et~al.} 2014, \aap, 563,
  A54

\bibitem[{{P{\^a}ris} {et~al.}(2017){P{\^a}ris}, {Petitjean}, {Ross}, {Myers},
  {Aubourg}, {Streblyanska}, {Bailey}, {Armengaud}, {Palanque-Delabrouille},
  {Y{\`e}che}, {Hamann}, {Strauss}, {Albareti}, {Bovy}, {Bizyaev}, {Niel
  Brandt}, {Brusa}, {Buchner}, {Comparat}, {Croft}, {Dwelly}, {Fan},
  {Font-Ribera}, {Ge}, {Georgakakis}, {Hall}, {Jiang}, {Kinemuchi},
  {Malanushenko}, {Malanushenko}, {McMahon}, {Menzel}, {Merloni}, {Nandra},
  {Noterdaeme}, {Oravetz}, {Pan}, {Pieri}, {Prada}, {Salvato}, {Schlegel},
  {Schneider}, {Simmons}, {Viel}, {Weinberg}, \& {Zhu}}]{Paris2016}
{P{\^a}ris}, I., {Petitjean}, P., {Ross}, N.~P., {et~al.} 2017, \aap, 597, A79

\bibitem[{{Peacock} \& {Dodds}(1996)}]{Peacock1996}
{Peacock}, J.~A., \& {Dodds}, S.~J. 1996, \mnras, 280, L19

\bibitem[{{Peebles}(1980)}]{Peebles1980}
{Peebles}, P.~J.~E. 1980, {The large-scale structure of the universe}

\bibitem[{{P{\'e}rez-R{\`a}fols} {et~al.}(2015){P{\'e}rez-R{\`a}fols},
  {Miralda-Escud{\'e}}, {Lundgren}, {Ge}, {Petitjean}, {Schneider}, {York}, \&
  {Weaver}}]{PerezRafols2014}
{P{\'e}rez-R{\`a}fols}, I., {Miralda-Escud{\'e}}, J., {Lundgren}, B., {et~al.}
  2015, \mnras, 447, 2784

\bibitem[{{P{\'e}rez-R{\`a}fols} {et~al.}(2018){P{\'e}rez-R{\`a}fols},
  {Font-Ribera}, {Miralda-Escud{\'e}}, {Blomqvist}, {Bird}, {Busca}, {du Mas
  des Bourboux}, {Mas-Ribas}, {Noterdaeme}, {Petitjean}, {Rich}, \&
  {Schneider}}]{Perezrafols2017}
{P{\'e}rez-R{\`a}fols}, I., {Font-Ribera}, A., {Miralda-Escud{\'e}}, J.,
  {et~al.} 2018, \mnras, 473, 3019

\bibitem[{{Pettini}(2004)}]{Pettini2003}
{Pettini}, M. 2004, in Cosmochemistry. The melting pot of the elements, ed.
  C.~{Esteban}, R.~{Garc{\'{\i}}a L{\'o}pez}, A.~{Herrero}, \&
  F.~{S{\'a}nchez}, 257--298

\bibitem[{{Pettini} {et~al.}(2001){Pettini}, {Ellison}, {Schaye}, {Songaila},
  {Steidel}, \& {Ferrara}}]{Pettini2001}
{Pettini}, M., {Ellison}, S.~L., {Schaye}, J., {et~al.} 2001, Astrophysics and
  Space Science Supplement, 277, 555

\bibitem[{Porciani \& Madau(2005)}]{Porciani2005}
Porciani, C., \& Madau, P. 2005, The Astrophysical Journal Letters, 625, L43

\bibitem[{{Pratt} {et~al.}(2017){Pratt}, {Stocke}, {Keeney}, \&
  {Danforth}}]{Pratt2017}
{Pratt}, C.~T., {Stocke}, J.~T., {Keeney}, B.~A., \& {Danforth}, C.~W. 2017,
  ArXiv e-prints, arXiv:1706.05103

\bibitem[{{Riemer-S{\o}rensen} {et~al.}(2015){Riemer-S{\o}rensen}, {Webb},
  {Crighton}, {Dumont}, {Ali}, {Kotu{\v s}}, {Bainbridge}, {Murphy}, \&
  {Carswell}}]{Signe2015}
{Riemer-S{\o}rensen}, S., {Webb}, J.~K., {Crighton}, N., {et~al.} 2015, \mnras,
  447, 2925

\bibitem[{{Savage} \& {Sembach}(1991)}]{Savage1991}
{Savage}, B.~D., \& {Sembach}, K.~R. 1991, \apj, 379, 245

\bibitem[{{Scannapieco} {et~al.}(2006){Scannapieco}, {Pichon}, {Aracil},
  {Petitjean}, {Thacker}, {Pogosyan}, {Bergeron}, \&
  {Couchman}}]{Scannapieco2006}
{Scannapieco}, E., {Pichon}, C., {Aracil}, B., {et~al.} 2006, \mnras, 365, 615

\bibitem[{{Seyffert} {et~al.}(2013){Seyffert}, {Cooksey}, {Simcoe}, {O'Meara},
  {Kao}, \& {Prochaska}}]{Seyffert2013}
{Seyffert}, E.~N., {Cooksey}, K.~L., {Simcoe}, R.~A., {et~al.} 2013, \apj, 779,
  161

\bibitem[{{Sheth} \& {Tormen}(1999)}]{Sheth1999}
{Sheth}, R.~K., \& {Tormen}, G. 1999, \mnras, 308, 119

\bibitem[{{Simcoe}(2011)}]{Simcoe2011}
{Simcoe}, R.~A. 2011, \apj, 738, 159

\bibitem[{{Smith} {et~al.}(2003){Smith}, {Peacock}, {Jenkins}, {White},
  {Frenk}, {Pearce}, {Thomas}, {Efstathiou}, \& {Couchman}}]{Smith2003}
{Smith}, R.~E., {Peacock}, J.~A., {Jenkins}, A., {et~al.} 2003, \mnras, 341,
  1311

\bibitem[{{Songaila} \& {Cowie}(1996)}]{Songaila1996}
{Songaila}, A., \& {Cowie}, L.~L. 1996, \aj, 112, 335

\bibitem[{{Tinker} {et~al.}(2010){Tinker}, {Robertson}, {Kravtsov}, {Klypin},
  {Warren}, {Yepes}, \& {Gottl{\"o}ber}}]{Tinker2010}
{Tinker}, J.~L., {Robertson}, B.~E., {Kravtsov}, A.~V., {et~al.} 2010, \apj,
  724, 878

\bibitem[{{Tytler} {et~al.}(1995){Tytler}, {Fan}, {Burles}, {Cottrell},
  {Davis}, {Kirkman}, \& {Zuo}}]{Tytler1995}
{Tytler}, D., {Fan}, X.-M., {Burles}, S., {et~al.} 1995, in QSO Absorption
  Lines, ed. G.~{Meylan}, 289

\bibitem[{{Vikas} {et~al.}(2013){Vikas}, {Wood-Vasey}, {Lundgren}, {Ross},
  {Myers}, {AlSayyad}, {York}, {Schneider}, {Brinkmann}, {Bizyaev},
  {Brewington}, {Ge}, {Malanushenko}, {Malanushenko}, {Muna}, {Oravetz}, {Pan},
  {P{\^a}ris}, {Petitjean}, {Snedden}, {Shelden}, {Simmons}, \&
  {Weaver}}]{Vikas2013}
{Vikas}, S., {Wood-Vasey}, W.~M., {Lundgren}, B., {et~al.} 2013, \apj, 768, 38

\bibitem[{{Vogt} {et~al.}(1994){Vogt}, {Allen}, {Bigelow}, {Bresee}, {Brown},
  {Cantrall}, {Conrad}, {Couture}, {Delaney}, {Epps}, {Hilyard}, {Hilyard},
  {Horn}, {Jern}, {Kanto}, {Keane}, {Kibrick}, {Lewis}, {Osborne},
  {Pardeilhan}, {Pfister}, {Ricketts}, {Robinson}, {Stover}, {Tucker}, {Ward},
  \& {Wei}}]{Vogt1994}
{Vogt}, S.~S., {Allen}, S.~L., {Bigelow}, B.~C., {et~al.} 1994, in \procspie,
  Vol. 2198, Instrumentation in Astronomy VIII, ed. D.~L. {Crawford} \& E.~R.
  {Craine}, 362

\bibitem[{{Wiersma} {et~al.}(2010){Wiersma}, {Schaye}, {Dalla Vecchia},
  {Booth}, {Theuns}, \& {Aguirre}}]{Wiersma2010}
{Wiersma}, R.~P.~C., {Schaye}, J., {Dalla Vecchia}, C., {et~al.} 2010, \mnras,
  409, 132

\bibitem[{{Wild} {et~al.}(2008){Wild}, {Kauffmann}, {White}, {York}, {Lehnert},
  {Heckman}, {Hall}, {Khare}, {Lundgren}, {Schneider}, \& {vanden
  Berk}}]{Wild2008}
{Wild}, V., {Kauffmann}, G., {White}, S., {et~al.} 2008, \mnras, 388, 227

\bibitem[{{Wolfe} {et~al.}(2005){Wolfe}, {Gawiser}, \& {Prochaska}}]{Wolfe2005}
{Wolfe}, A.~M., {Gawiser}, E., \& {Prochaska}, J.~X. 2005, \araa, 43, 861

\bibitem[{{Zehavi} {et~al.}(2005){Zehavi}, {Zheng}, {Weinberg}, {Frieman},
  {Berlind}, {Blanton}, {Scoccimarro}, {Sheth}, {Strauss}, {Kayo}, {Suto},
  {Fukugita}, {Nakamura}, {Bahcall}, {Brinkmann}, {Gunn}, {Hennessy},
  {Ivezi{\'c}}, {Knapp}, {Loveday}, {Meiksin}, {Schlegel}, {Schneider},
  {Szapudi}, {Tegmark}, {Vogeley}, {York}, \& {SDSS
  Collaboration}}]{Zehavi2005}
{Zehavi}, I., {Zheng}, Z., {Weinberg}, D.~H., {et~al.} 2005, \apj, 630, 1

\end{thebibliography}

\end{document}